\documentclass[12pt,letterpaper]{article}
\usepackage{jheppub}
\usepackage{amsmath, amssymb}
\usepackage{verbatim}

\usepackage{graphicx}
\usepackage{subfigure}
\input{epsf}
\usepackage{epsfig}
\usepackage{epstopdf}
\usepackage{amsthm}
\usepackage{xcolor}
\usepackage{slashed}

\def\({\left(} \def\){\right)}
\def\[{\left[} \def\]{\right]}

\def\del{{\partial}}

\def\f{\frac}
\def\lan{\langle}
\def\ran{\rangle}

\def\l{\lambda}
\def\s{\sigma}

\def\eps{\epsilon}

\def\mo{\mathcal{O}}

\def\md{\mathcal{D}}

\def\nn{\nonumber}
\def\lwf{\l_\text{\tiny{WF}}}
\def\gwf{g_\text{\tiny{WF}}}
\def\ygn{\eta_\text{\tiny{GNY}}}
\def\lgn{\l_\text{\tiny{GNY}}}
\def\ggn{g_\text{\tiny{GNY}}}
\def\yuk{\eta}

\title{The defect $b$-theorem under bulk RG flows}
\author{Tom Shachar$^a$,}
\author{Ritam Sinha$^b$, and}
\author{Michael Smolkin$^a$}

\affiliation[a]{The Racah Institute of Physics, The Hebrew University of Jerusalem, \\ Jerusalem 91904, Israel }
\affiliation[b]{Department of Mathematics, King’s College London, Strand, London, WC2R 2LS, UK\\}

\emailAdd{tom.shachar@mail.huji.ac.il}
\emailAdd{ritam.sinha@kcl.ac.uk}
\emailAdd{michael.smolkin@mail.huji.ac.il}

\abstract{It is known that for RG flows confined to a two-dimensional defect, where the bulk maintains its conformal nature, the coefficient of the Euler density in the defect's Weyl anomaly (termed $b$) cannot increase as the flow progresses from the ultraviolet to the infrared, a principle known as the $b$-theorem. In this paper, we investigate whether this theorem still holds when the bulk, instead of being critical, also undergoes an RG flow. To address this question, we examine two distinct and perturbatively tractable examples. Our analysis reveals that a straightforward extension of the $b$-theorem to these cases of RG flows fails.}

\begin{document}
\maketitle
\section{Introduction}

The renormalization group (RG) flow serves as a foundational framework for identifying the degrees of freedom governing low-energy phenomena. Its core premise lies in simplifying theories by disregarding their microscopic details while preserving their low-energy physics. This simplification inevitably reduces the number of degrees of freedom, sparking a longstanding debate regarding the quantification of this reduction. Zamolodchikov's c-theorem \cite{Zamolodchikov:1986gt} provided the first precise quantification of this kind for a broad class of two-dimensional quantum field theories, catalyzing numerous advancements across various spacetime dimensions and expanding our understanding of RG flows and their implications.

Extending Zamolodchikov's theorem to higher dimensions, let alone to the case of QFTs with defects, has proven to be a challenging endeavour, leading to ongoing research efforts aimed at elucidating the nature of RG flows beyond the two-dimensional case \cite{Cardy:1988cwa,Cappelli:1990yc,Osborn:1991gm,Myers:2010tj,Jafferis:2011zi,Komargodski:2011vj,Komargodski:2011xv,Casini:2012ei,Elvang:2012st,Yonekura:2012kb,Grinstein:2013cka,Jack:2013sha,Giombi:2014xxa,Jack:2015tka,Cordova:2015fha,Casini:2015woa,Casini:2017vbe,Fluder:2020pym,Delacretaz:2021ufg}.

In this paper, we delve into the study of RG flows in the presence of two-dimensional defects. Our focus lies exclusively on scenarios where the bulk QFT is a $d$-dimensional Euclidean field theory, with the state being the flat space vacuum state. In such configurations, both the defect and the bulk can undergo RG flows, thereby rendering existing analogs of the c-theorem inapplicable. Despite the extensive history of defects in both two and higher dimensions \cite{Cardy:1984bb,Cardy:1989ir,McAvity:1995zd,Affleck:1995ge,Sachdev99,Vojta_2000,Polchinski:2011im,Gaiotto:2013nva,Billo:2016cpy,Bianchi:2015liz,Solodukhin:2015eca,Fursaev:2016inw,Lauria:2018klo,Gadde:2016fbj,Herzog:2020bqw,Giombi:2021uae,Liu_2021,Herzog:2022jqv}, a robust candidate to quantify the reduction of degrees of freedom when both the bulk and the defect undergo simultaneous RG flows remains elusive and rarely addressed \cite{Herzog:2019rke, Bianchi:2019umv, Herzog:2021hri}.

In contrast, defect RG flows with fixed conformal bulk, also known as DRG in the literature, have undergone extensive studies \cite{Affleck:1991tk,Yamaguchi:2002pa,Azeyanagi:2007qj,Estes:2014hka,Andrei:2018die,Kobayashi:2018lil,Casini:2018nym,Lauria:2020emq,Giombi:2020rmc,Wang:2020xkc,Nishioka:2021uef,Sato:2021eqo,CarrenoBolla:2023vrv}.\footnote{Recent examples and perturbative calculations concerning defects span a wide array of systems and models, as detailed in \cite{Padayasi:2021sik,Rodriguez-Gomez:2022gbz,Cuomo:2021kfm,Rodriguez-Gomez:2022gif,Castiglioni:2022yes,Krishnan:2023cff,Cuomo:2023qvp,Drukker:2023jxp,Shimamori:2024yms}.} Numerous exact results regarding RG flows on line defects \cite{Friedan:2003yc,Casini:2016fgb,Cuomo:2021rkm,Casini:2022bsu} and their higher-dimensional generalizations \cite{Jensen:2015swa,Wang:2021mdq,Shachar:2022fqk} have been established. In particular, the so-called $b$-theorem \cite{Jensen:2015swa,Shachar:2022fqk} asserts that the dimensionless "central charge," which multiplies the Euler density in the defect's Weyl anomaly, necessarily decreases or remains constant along the DRG flow from the UV to an IR fixed point, providing a quantitative diagnostic for the irreversibility of the DRG flows on two-dimensional defects.

In this work, we pose a natural question: does the $b$-theorem still hold if the condition of a conformal bulk is relaxed, subjecting it to changes under an RG flow?
We term this version of the $b$-theorem a naive generalization, and proceed to test it in two specific examples in $4-\epsilon$ dimensions: the $O(N)$ vector model and the Gross-Neveu-Yukawa model. Our calculations illustrate that when the flow extends beyond the defect, implying that the bulk is not conformally invariant and undergoes changes along the RG flow, the naive extension of the $b$-theorem is compromised. 

Previous works in this direction either study exactly marginal deformations in the bulk \cite{Herzog:2019rke, Bianchi:2019umv, Herzog:2021hri}, which correspond to spanning different theories along the conformal manifold, or rely on the assumption that the bulk deformation does not induce a defect RG flow up to a sufficiently high order in the weakly relevant bulk deformation -  protected defects \cite{Green:2007wr,Sato:2020upl}. However, to the best of our knowledge, explicit quantum field theory examples of higher dimensional protected defects do not exist in the literature.

The paper is organized as follows: in Section \ref{dRGgen}, we employ conformal perturbation theory to derive perturbative beta functions for a generic CFT deformed by a set of relevant operators in the bulk and on the two-dimensional defect. Sections \ref{vector_model} and \ref{GNY} are dedicated to scrutinizing RG flows in the presence of defects, using two examples of QFTs in $4-\epsilon$ dimensions that are perturbatively tractable: the $O(N)$ vector model and the Gross-Neveu-Yukawa model. We conclude with a discussion in Section \ref{discussion}, followed by two appendices. Appendix \ref{F-Appx} presents calculations of the free energy for critical models with defects studied in this paper to explicitly evaluate the numerical values of the $b$-anomaly. In Appendix \ref{GNY-RG}, we derive the full set of beta functions for the critical Gross-Neveu-Yukawa model with a two-dimensional defect.

\section{RG flow in the presence of 2D defect}
\label{dRGgen}

In this section we derive perturbative beta functions in the presence of two-dimensional defect. In the next section, we use these results to analyze the effect of bulk RG flows on the defect $b$-theorem. 
	
Consider a $d$-dimensional Euclidean defect CFT with a set of local operators $\mathcal{O}_i$ and $\mathcal{O}_\alpha$, each having scaling dimensions $\Delta_i=2-\epsilon$ and $\Delta_\alpha=d-\epsilon$, respectively. For simplicity, we assume that the 2-point functions of these operators have unit amplitude. Their OPE coefficients are denoted as follows
\begin{eqnarray}
 \mo_i(x) \mo_j(0) \sim  { C_{ij}^k\over |x|^{\Delta_i+\Delta_j-\Delta_k}} \mo_k(0)  + \ldots ~,
 \nonumber
 \\
 \mo_\alpha(x) \mo_\beta(0) \sim { C_{\alpha \beta}^{\nu} \over |x|^{\Delta_\alpha+\Delta_\beta-\Delta_\nu}} \mo_\nu(0)  + \ldots ~,
  \nonumber
 \\
  \mo_i(0)  \mo_\alpha(x) \sim { C_{i\alpha}^j \over |x|^{\Delta_\alpha+\Delta_i-\Delta_j}}  \mo_j(0)  + \ldots ~,
 \end{eqnarray}

We use these operators to initiate an RG flow in both the bulk and the two-dimensional defect by deforming the CFT as follows
\begin{align}
 S=S_\text{CFT} +\delta S = S_\text{CFT} + g^i_0 \int d^2 \sigma \sqrt{\gamma} \, \mo_i + \lambda^\alpha_0 \int  \,d^dx~\mo_\alpha ~,
\end{align}
where $\left\lbrace\sigma_a \right\rbrace_{a=1,2}$ and $\left\lbrace x^\mu \right\rbrace_{\mu=1,\ldots,d}$ denote the coordinates that parametrize the two-dimensional defect and the bulk, respectively, while $\gamma_{ab}$ represents the induced metric. Assuming that $\epsilon\ll 1$, operators with Greek indices represent weakly relevant deformations in the bulk, while operators with Latin indices correspond to weakly relevant deformations in the two-dimensional defect. By construction, the CFT without a defect and with the bulk coupling $g_0^i$ tuned to zero, serves as a UV fixed point of the flow, whereas the IR end of the flow is located in its vicinity. Hence, the couplings remain weak throughout the flow, and one can perturbatively expand  around the UV CFT, 
 \begin{align}
 e^{-\delta S} = 1 - \delta S + {\lambda^\alpha_0 \lambda^\beta_0 \over 2} \int_\mathcal{B}  \int_\mathcal{B} \mo_\alpha \mo_\beta +  {g^i_0 g^j_0 \over 2} \int_\mathcal{D} \mo_i \mo_j 
 +  g^i_0 \lambda^\alpha_0 \int_\mathcal{D}\int_\mathcal{B} \mo_i \mo_\alpha ~.
\end{align}
where we introduced a shorthand notation
\begin{equation}
 \int_\mathcal{D} = \int d^2\sigma \sqrt{\gamma} ~, \quad \int_\mathcal{B}=\int d^dx ~.
\end{equation}

To obtain the defect and the bulk at a specific scale $\mu$, we integrate out distances within the range $0 \leq \ell \leq \mu^{-1}$. This calculation essentially involves excluding a small ball with a radius of $\mu^{-1}$ around the operators on the right-hand side of the expression above. The resulting relation between the dimensionless couplings at scale $\mu$ and their bare counterparts is given by
\begin{eqnarray}
 \lambda^\nu\mu^\epsilon&=&\lambda^\nu_0- {\pi^{d\over 2} \over \Gamma\big({d\over 2}\big)}   { \mu^{-\epsilon}\over \epsilon} \, C ^\nu_{\alpha\beta} \, \lambda_0^\alpha \, \lambda_0^\beta + \ldots ~,
 \nonumber \\
 g^i\mu^\epsilon&=&g^i_0 - \pi C^i_{jk} \, g_0^j \, g_0^k  \, { \mu^{-\epsilon}\over \epsilon}  
 - {2\pi^{d\over 2} \over \Gamma\big({d\over 2}\big)}   \, { \mu^{-\epsilon}\over \epsilon} \, C ^i_{j \alpha} \, g^j_0\, \lambda_0^\alpha \, + \ldots~.
 \label{bareVSren}
 \end{eqnarray}
Hence, the flow of the couplings follows the pattern
\begin{eqnarray}
 \beta^i &=& \mu {d g^i\over d\mu} = -\epsilon g^i + \pi C^i_{jk} g^j g^k + {2\pi^{d\over 2} \over \Gamma\big({d\over 2}\big)} C^i_{j \alpha} \, g^j  \lambda^\alpha  
 + \mathcal{O}(g^3)~,
 \nonumber \\
 \beta^\nu &=& \mu {d \lambda^\nu \over d\mu} = -\epsilon \lambda^\nu + {\pi^{d\over 2} \over \Gamma\big({d\over 2}\big)} C^\nu_{\alpha \beta} \lambda^\alpha \lambda^\beta + \mathcal{O}(\lambda^3)~.
 \label{beta_functions}
 \end{eqnarray}

\section{$O(N)$ vector model with a defect}
\label{vector_model}

In this section, we delve into a specific example of the general setup introduced in the previous section. We focus on evaluating the fixed points within the two-dimensional coupling space and analyze the RG flows connecting them. Our examination reveals that when the flow extends beyond the defect, implying that the bulk is not conformally invariant and undergoes changes along the RG flow, the naive extension of the $b$-theorem is compromised.

Consider a free massless $O(N)$ vector field $\phi$ in $d=4-\epsilon$ dimensions. This CFT can be perturbed by a weakly relevant $O(N)$ symmetric quartic interaction in the bulk and an $O(N)$ symmetric quadratic interaction on a two-dimensional spherical defect,
	\begin{align} \label{ONaction}
		S= \int_\mathcal{B} ~ \Big(\f{1}{2}(\partial\phi)^2+ \l \mu^{\eps} \phi^4 \Big) + g \mu^{\eps} \int_{\mathcal{S}^2}~\phi^2 ~,
	\end{align}
Neither the two-point function of $\phi^4$ nor that of $\phi^2$ is normalized.\footnote{For brevity, we suppress the flavor index of the vector field $\phi$, {\it e.g.,} $\phi^2=\sum_{a=1}^N \phi_a \phi_a $, and similarly for $\phi^4=(\phi^2)^2$. The exact expressions for the two-point functions of the free theory are as follows,
\begin{equation}
  \langle \phi_a(x) \phi_c(0)\rangle= { \delta_{ac} \, C_\phi \over |x|^{d-2}} ~, \quad
 \langle \phi^2(x) \phi^2(0)\rangle= {2N C_\phi^2 \over |x|^{2(d-2)}} ~, \quad  \langle \phi^4 (x) \phi^4(0)\rangle= {8N(N + 2) C_\phi^4 \over |x|^{4(d-2)}} ~.
\end{equation}
}
However, for the purpose of using (\ref{beta_functions}), we only need to consider the relevant OPE coefficients, which can be obtained using Wick contractions,
\begin{eqnarray}
 \phi^2(x) \phi^2(0) &=& {C_\mathcal{D}  \over |x|^{d-2}} ~ \phi^2 (0) + \ldots ~, 
\nonumber \\
\phi^2(x) \phi^4(0) &=& {C_\mathcal{BD} \over |x|^{2(d-2)}}  ~\phi^2 (0) + \ldots  ~,
 \\
 \phi^4(x) \phi^4(0) &=& {C_\mathcal{B}  \over |x|^{2(d-2)}} ~ \phi^4 (0) + \ldots ~,
 \nonumber
\end{eqnarray}
with
	\begin{equation} \label{freeOPE}
		C_\mathcal{D} = 4C_\phi\, , ~~~~C_\mathcal{BD} = 4(N+2)C_{\phi}^2\, , ~~~ C_\mathcal{B} = 8(N+8)C_{\phi}^2\, , ~~~ C_\phi={\Gamma\({d-2\over 2}\)\over 4\pi^{d\over 2}} ~.
	\end{equation}	
Hence, the beta functions in this model take the form,
	\begin{eqnarray}
		\beta_\l &=& -\eps\l + \pi^2 C_\mathcal{B} \l^2 ~,
		\nonumber \\
	 \label{betaDWF} 
		\beta_g &=& -\eps g + \pi C_\mathcal{D} g^2 + 2\pi^2 C_\mathcal{BD} \l g ~.
	\end{eqnarray}
In total, there are four fixed points of the flow: 

\begin{equation}
	(g_*, \l_*)=\{(0,0),(0, \lwf),(g_\text{\tiny GD}, 0),(\gwf,\lwf)\}~~,
\end{equation}
where
	\begin{equation} \label{criticalWF}
		\lwf=\f{2\pi^2\, \eps}{N+8} ~ ,\quad g_\text{\tiny G}=\epsilon \pi~, \quad \gwf=\f{6\pi\eps}{N+8}~.
	\end{equation}
The first two fixed points with $g_*=0$ correspond to the Gaussian and Wilson-Fisher (WF) CFTs without defects. The two remaining fixed points in the two-dimensional $(g, \lambda)$-plane represent DCFTs with non-vanishing defects. For example, the point $(g_\text{\tiny G},0)$ describes a non-trivial defect in a free field theory \cite{Shachar:2022fqk}, which we will refer to as the Gaussian DCFT (dG). Similarly, $(\gwf,\lwf)$ describes a non-trivial conformal defect embedded into the WF CFT bulk \cite{Trepanier2023,Giombi2023,RavivMoshe2023}, which we will refer to as dWF for brevity.

\begin{figure}[h!]
		\centering
		\includegraphics[width=0.45\columnwidth]{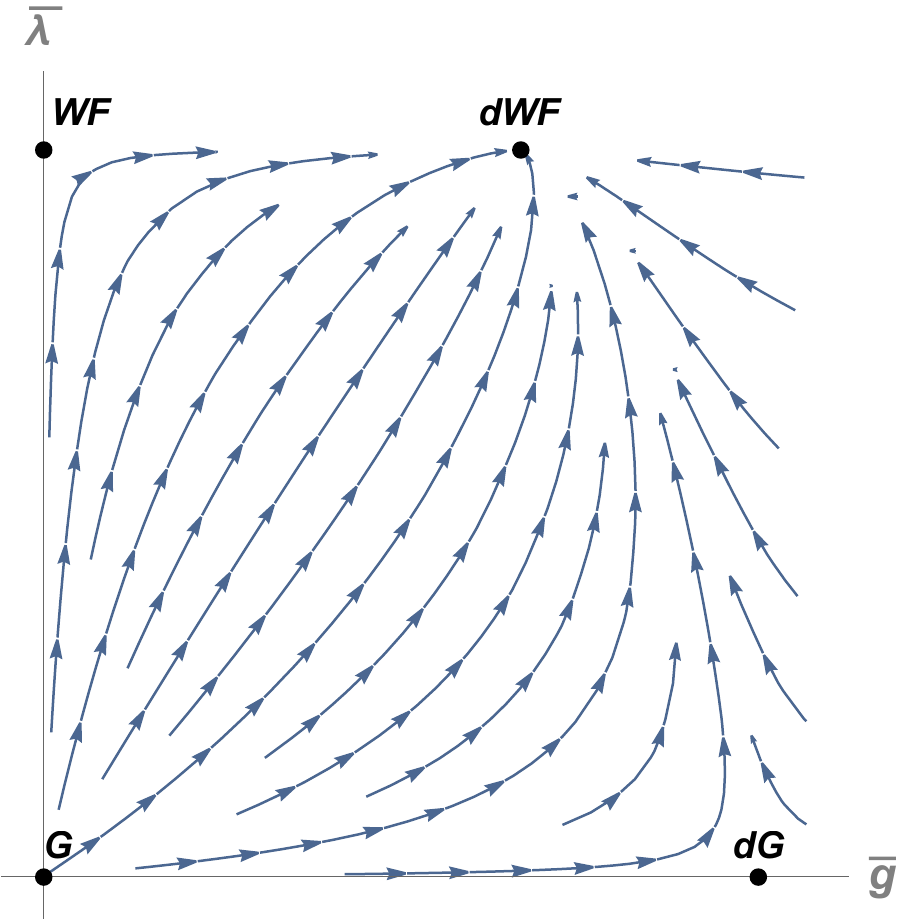}
		\caption{RG flow trajectories of (\ref{betaDWF}) are shown for $N=\epsilon=1$. Here, $\bar\lambda=\lambda/\lwf$ and $\bar g=g/\gwf$.  Arrows indicate the direction towards the IR. The bold point at the origin (G) represents the infrared repulsive massless free vector model without defect. The Wilson-Fisher (WF) conformal field theory without defect and the Gaussian DCFT (dG) are denoted by solid dots on the $\bar\lambda$ and $\bar g$ axes, respectively. The solid dot denoted by dWF corresponds to the Wilson-Fisher fixed point with defect, ($\gwf,\lwf$), which serves as an IR attractor of the RG flows in the two-dimensional coupling space.}
		\label{Fig1}
	\end{figure}

By analyzing the matrix of derivatives of the beta functions in (\ref{betaDWF}), one can show that the trivial defect in the Gaussian and WF CFTs is infrared repulsive under the $\phi^2$ deformation. Additionally, the dG has an unstable direction triggered by turning on the relevant $\phi^4$ deformation, which causes the dG to flow towards dWF, as shown in Fig.\ref{Fig1}. In the remaining part of this section, we demonstrate that along this RG flow, the $b$-anomaly increases, thus violating the naive generalization of the $b$-theorem for 2D surface defects in the presence of bulk RG flows.

To determine the value of the anomaly constant $b$ for dWF, we examine the RG trajectory of (\ref{betaDWF}) with $\lambda=\lwf$. In this scenario, $\beta_\lambda=0$ throughout the flow, indicating that the bulk remains intact and is represented by the WF CFT. In contrast, the defect changes: it is trivial in the UV but is characterized by $\gwf\neq 0$ in the IR. This RG trajectory corresponds to a horizontal line connecting the WF fixed point on the $\lambda$-axis with dWF in Fig. \ref{Fig1}. Therefore, for this specific RG trajectory, we can use the general result for the increment of $b$, as obtained in \cite{Shachar:2022fqk} through conformal perturbation theory. In our context, it takes the form,
\begin{equation}
b_\text{\tiny{WF}}=- {N\over 8} \bar\epsilon^3 + \mathcal{O}(\bar\epsilon^4)~,
\label{b_WF}
\end{equation}
where $\bar\epsilon=2-\Delta_{\phi^2}$, and $\Delta_{\phi^2}$ represents the scaling dimension of $\phi^2$ at the Wilson-Fisher fixed point, {\it i.e.,}
\begin{equation}
		\bar{\eps} = -{\del\beta_g\over \del g}\bigg|_{g=0,~\lambda=\lwf}= {6\, \eps \over N+8}~.
\end{equation}
In Appendix \ref{F-Appx}, we perform a direct calculation of the partition function to independently verify the above value of $b_\text{\tiny{WF}}$. Similarly \cite{Shachar:2022fqk},
\begin{equation}
b_\text{\tiny{G}}=- {N\over 8} \epsilon^3 + \mathcal{O}(\epsilon^4)~.
\end{equation}
Hence, for all $N$, we obtain 
	\begin{equation}
		b_\text{\tiny{WF}}-b_{\text{\tiny{G}}}=\f{N\left( \eps^3-\bar{\eps}^3\right) }{8} >0 ~,
	\end{equation}
leading to a violation of the naive generalization of the $b$-theorem under the RG flow from the dG to dWF.


	
\section{Gross-Neveu-Yukawa model with a defect}
\label{GNY}

Consider now the same two-dimensional spherical defect as in the previous section, but this time embedded in the $d=4-\epsilon$ dimensional bulk described by a single bosonic field, $\phi$, coupled to $N_\text{F}$ Dirac fields,
\begin{align}
S=  \int_\mathcal{B} ~ \Big(\bar{\psi}\slashed{\partial}\psi +\f{1}{2}(\partial\phi)^2+\yuk\,\mu^{\eps\over 2}\bar{\psi}\psi\phi+\f{\l \mu^{\eps}}{4!}\phi^4\Big) + g \mu^{\eps}\int_{\mathcal{S}^2}\phi^2 ~.
 \label{GNYaction}
\end{align}
This action describes the $U(N_\text{\tiny F})$ invariant Gross-Neveu-Yukawa (GNY) model with a two-dimensional spherical defect. To maintain brevity, we suppress the flavor index of the vector field $\psi$. 

The RG flow in this model occurs in the three-dimensional space of couplings $(\eta,\lambda, g)$. The associated beta functions are derived in Appendix \ref{GNY-RG}, while here we present the final expressions
\begin{eqnarray}
		\beta_\l &=& -\eps\l +\f{1}{8\pi^2}\Big(\,\f{3}{2}\l^2+N\l \yuk^2-6N\yuk^4\Big)+ \ldots ~, \nonumber \\
		\beta_{\yuk^2} &=& -\eps \yuk^2+\f{N+6}{16\pi^2}\yuk^4 + \ldots ~, \label{GNY-beta}\\
		\beta_g &=&  -\eps g + \f{g^2}{\pi} + \f{g}{16\pi^2}\left(\l+N\yuk^2\right) + \ldots ~,
		\nonumber
\end{eqnarray}
where ellipses denote higher-order corrections in the small coupling constants. Here, $N=\text{Tr}(\mathbb{I}_{4\times 4}) N_\text{\tiny F} = 4 N_\text{\tiny F}$ represents the total number of fermionic degrees of freedom in four dimensions. Notably, the RG flow equations (\ref{GNY-beta}) cannot be recovered using (\ref{beta_functions}) because conformal perturbation theory in Section \ref{vector_model} is restricted to quadratic order in the coupling constants, where the knowledge of OPE coefficients is enough to recover the structure of the beta functions. In contrast, for consistency, the beta functions (\ref{GNY-beta}) must necessarily include quartic corrections in $\eta$.

The two-dimensional plane $\eta=0$ encompasses all the fixed points previously discussed in section \ref{vector_model}. In addition to these, there exist two new fixed points: $(\yuk_*,\l_*,g_*)=(\ygn,\lgn,\ggn)$ and $(\ygn,\lgn,0)$, with and without defect, respectively, where
\begin{eqnarray}
	\ygn&=&4\pi\sqrt{\f{\eps}{N+6}} ~, \nonumber \\
	\lgn&=&\f{384\pi^2 N\eps}{\left(N+6\right)\left(N-6+\sqrt{N^2+132N+36}\right)}~, \label{ggny} \\
	\ggn&=&\pi\eps\bigg(1-\f{6+5N+\sqrt{N^2+132N+36}}{6\left(N+6\right) } \,\bigg) ~.
	\nonumber 
\end{eqnarray}

\begin{figure}[h!]
		\centering
		\includegraphics[width=0.45\columnwidth]{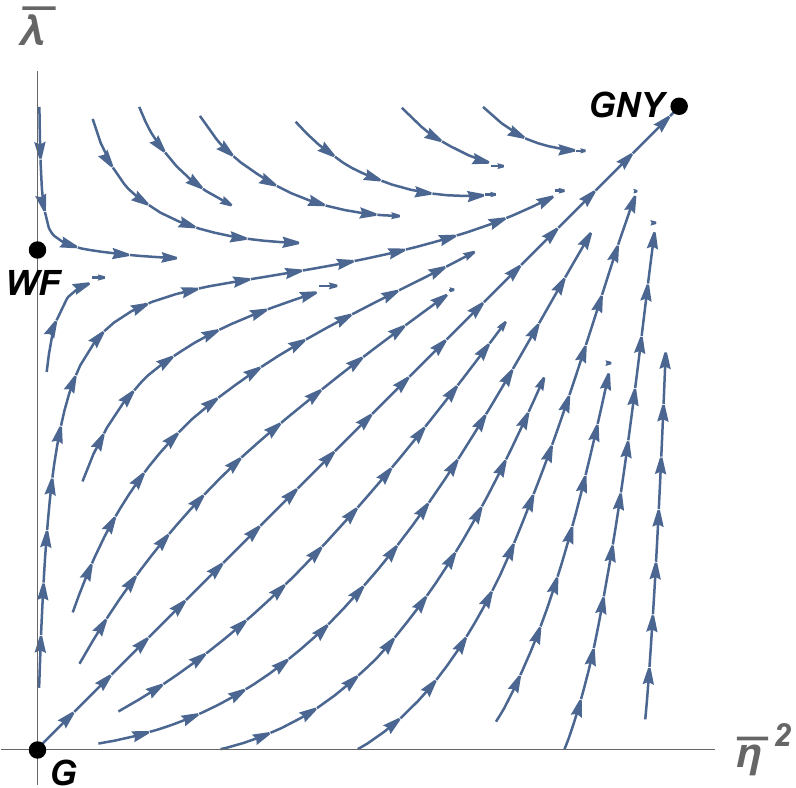}
		\caption{The plot depicts the $g=0$ plane accommodating RG flow trajectories without a defect, obtained from (\ref{ggny}) for $N=\epsilon=1$. Here, $\bar\lambda=\lambda/\lgn$ and $\bar\eta=\eta/\ygn$. Arrows indicate the direction towards the IR. The bold point at the origin represents the infrared repulsive Gaussian CFT (G) with a massless free scalar and $N_\text{\tiny F}$ massless Dirac fields. The solid dots on the $\bar\lambda$ axis and in the upper right corner denote the critical Wilson-Fisher (WF) and Gross-Neveu-Yukawa (GNY) models, respectively.}
		\label{fig:GNY_RG}
	\end{figure}

The fixed point $(\ygn,\lgn,0)$ belongs to the $g=0$ plane accommodating all possible RG flows without a defect. As shown in Fig. \ref{fig:GNY_RG}, this fixed point is an IR stable attractor of the RG trajectories restricted to this plane.

\begin{figure}[h!]
		\centering
		\includegraphics[width=0.45\columnwidth]{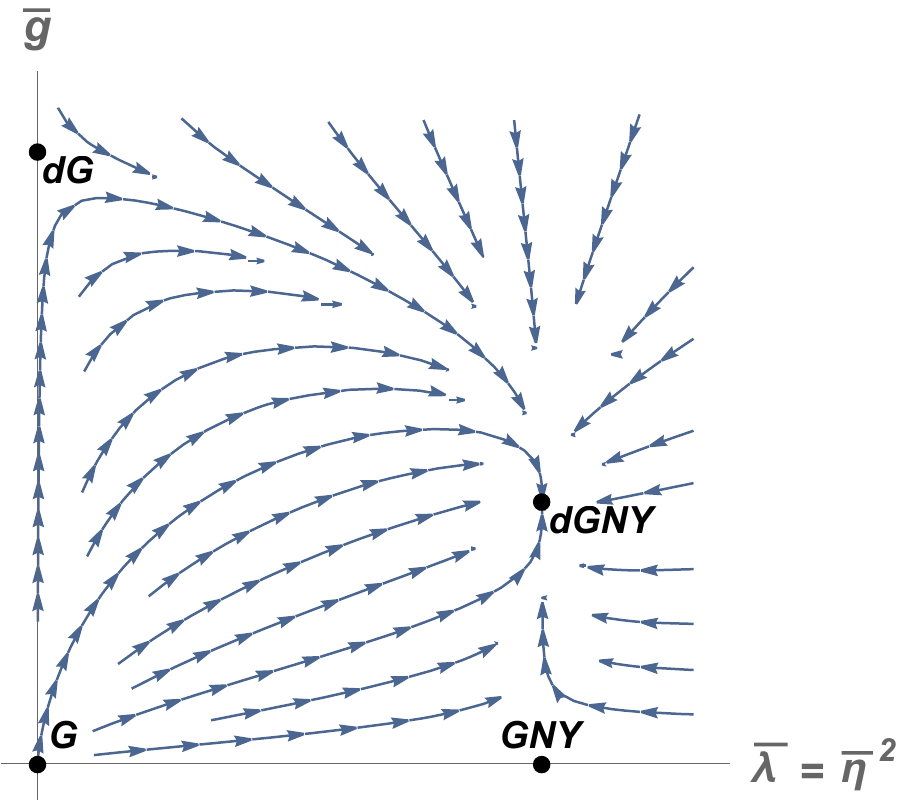}\quad\quad \includegraphics[width=0.45\columnwidth]{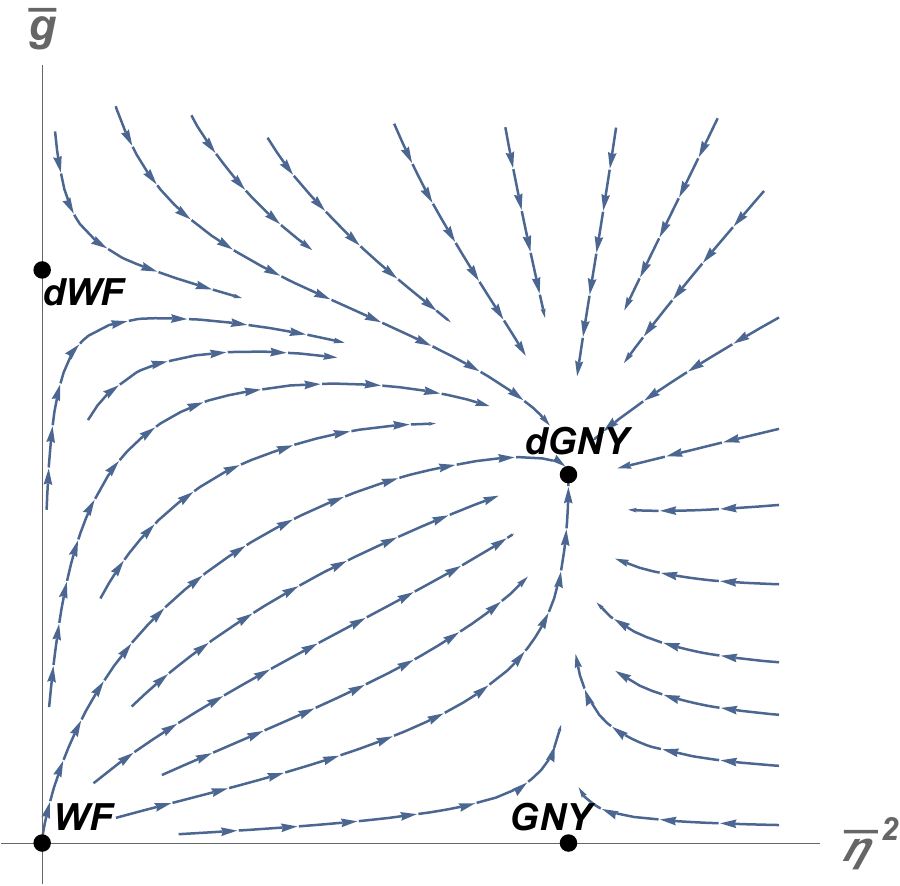}
		\caption{The plots display planar sections of RG flow trajectories obtained from (\ref{ggny}) for $N=\epsilon=1$. Here, $\bar\lambda=\lambda/\lgn$, $\bar\eta=\eta/\ygn$, and $\bar g=g/\ggn$. Arrows indicate the direction towards the IR. The solid dot denoted by dGNY corresponds to the critical Gross-Neveu-Yukawa (GNY) model with a defect, $(\ygn,\lgn,\ggn)$, serving as an IR attractor of the RG flows in the three-dimensional coupling space. The critical GNY model without a defect is represented by a solid dot on the horizontal axis. \underline{{\bf Left panel:}} Lines of RG flow are displayed in the $\bar\lambda=\bar\eta^2$ plane. The origin represents the infrared repulsive Gaussian CFT (G) with a massless free scalar and $N_\text{\tiny F}$ massless Dirac fields. The solid dot on the $\bar g$ axis denotes a Gaussian CFT with a non-trivial defect (dG). \underline{{\bf Right panel:}} Lines of RG flow are displayed in the $\bar\lambda=\bar\eta^2+\lwf/\lgn(1- \bar\eta^2)$ plane. The origin represents the infrared repulsive critical Wilson-Fisher (WF) fixed point without a defect. The solid dot on the $\bar g$ axis corresponds to WF with a non-trivial conformal defect (dWF).}
	\label{fig:dGNY_RG}
	\end{figure}

Analysis of the $3\times 3$ matrix of derivatives of the beta functions (\ref{GNY-beta}) leads to the conclusion that the previously IR attractive fixed point $(0,\lwf,\gwf)$ becomes unstable under Yukawa deformation. Similarly, the critical GNY model without a defect becomes IR repulsive when introducing the $\phi^2$ deformation onto the two-dimensional defect surface. Furthermore, as depicted in Fig. \ref{fig:dGNY_RG}, the DCFT characterized by $(\ygn,\lgn,\ggn)$, denoted by dGNY in the figure, stands as the only IR stable fixed point in the three-dimensional coupling space of the GNY model with the type of defect studied in this section.

The $b$-anomaly of the IR stable DCFT can be derived by considering the RG flow originating from $(\ygn,\lgn,0)$ and terminating at $(\ygn,\lgn,\ggn)$ with a fixed GNY bulk, $\yuk=\ygn$ and $\l=\lgn$, along the RG trajectory. This flow corresponds to a line parallel to the $\bar{g}$-axis in Fig. \ref{fig:dGNY_RG}, connecting the critical GNY model without a defect with dGNY. Using the general result for the increment of $b$, as obtained in \cite{Shachar:2022fqk} through conformal perturbation theory, yields
\begin{equation}
b_\text{\tiny{GNY}}= - { \bar\epsilon^3\over 8} + \mathcal{O}(\bar\epsilon^4)~,
\label{b_GNY}
\end{equation}
where $\bar\epsilon=2-\Delta_{\phi^2}$, and $\Delta_{\phi^2}$ represents the scaling dimension of $\phi^2$ in the critical GNY model,
\begin{equation}
		\bar{\eps} = -{\del\beta_g\over \del g}\bigg|_{(\ygn,\lgn,0)}=  \eps\bigg(1-\f{6+5N+\sqrt{N^2+132N+36}}{6\left(N+6\right) } \,\bigg) ~.
\end{equation}
In Appendix \ref{F-Appx}, we independently verify the above value for $b_\text{\tiny{GNY}}$ through a direct calculation of the partition function.
	
For $N=0$, the anomaly takes the value of $b_\text{\tiny{GNY}}|_{N=0}=-\frac{\epsilon^3}{27}$, which, as expected, matches $b\text{\tiny{WF}}$ in (\ref{b_WF}) for a single boson. It then increases with $N$ and vanishes at $N=12$. Consequently, for $N<12$ the naive generalization of the $b$-theorem is violated for two distinct RG flows that end at $(\ygn,\lgn,\ggn)$. The first scenario where it is violated, occurs when the flow starts at dG, leading to an increase in the value of $b$ for all $N<12$, see left panel of Fig. \ref{fig:dGNY_RG}. In the second scenario, the anomaly increases for any strictly positive $N<12$ when the flow starts at dWF and terminates at dGNY, see right panel of Fig. \ref{fig:dGNY_RG}.

\begin{figure}[h!]
		\centering
		\includegraphics[width=0.47\columnwidth]{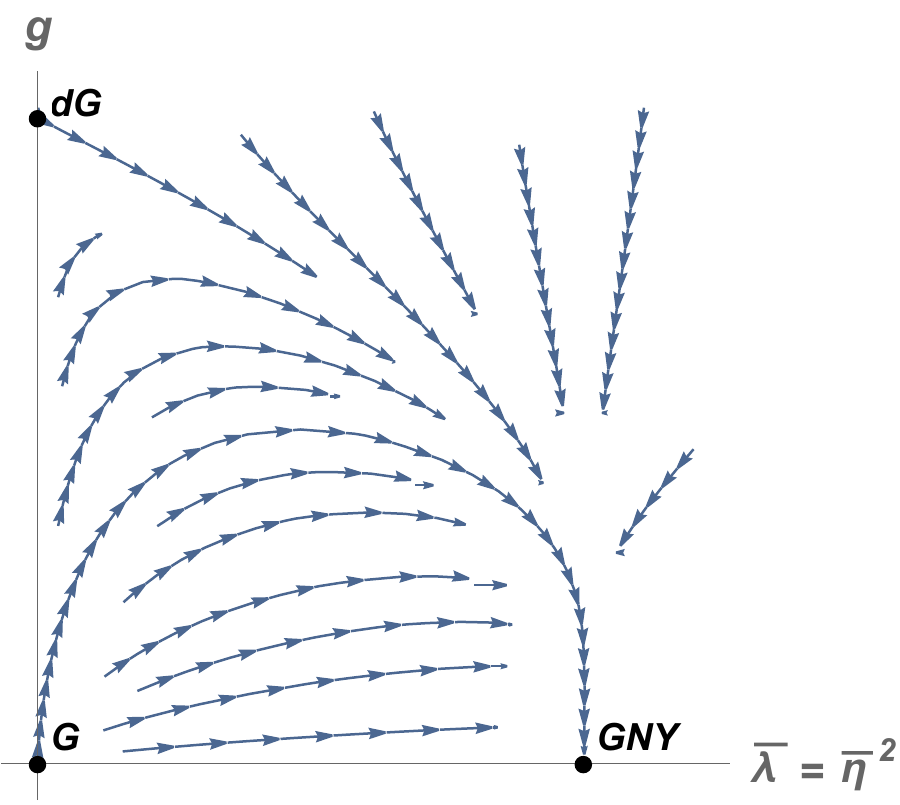}\quad\quad \includegraphics[width=0.45\columnwidth]{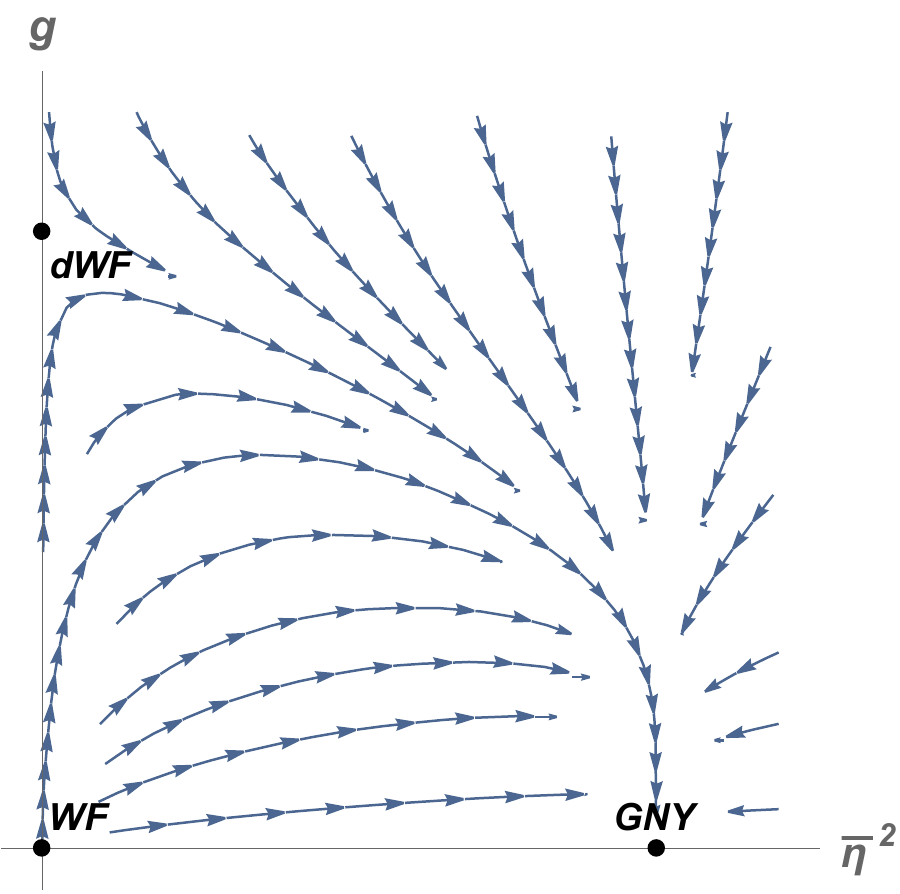}
		\caption{The plots depict planar sections of RG flow trajectories, obtained from (\ref{ggny}) for $N=12$ and $\epsilon=1$. Here, $\bar\lambda=\lambda/\lgn$ and $\bar\eta=\eta/\ygn$. Arrows indicate the direction towards the IR. The critical GNY model without a defect is represented by a solid dot on the horizontal axis. For $N=12$, this fixed point is an IR stable attractor of the RG flows in the three-dimensional coupling space.  \underline{{\bf Left panel:}} Lines of RG flow are displayed in the $\bar\lambda=\bar\eta^2$ plane. The origin represents the infrared repulsive Gaussian CFT (G) with a massless free scalar and $N_\text{\tiny F}$ massless Dirac fields. The solid dot on the $g$ axis denotes a Gaussian CFT with a non-trivial defect (dG). \underline{{\bf Right panel:}} Lines of RG flow are displayed in the $\bar\lambda=\lwf/\lgn=1$ plane. The origin represents the infrared repulsive critical Wilson-Fisher (WF) fixed point without a defect. The solid dot on the $g$ axis corresponds to WF with a non-trivial conformal defect (dWF).}
	\label{fig:dGNY_RG12}
	\end{figure}

Moreover, as $\ggn$ vanishes in the limit $N\to 12$, the fixed point dGNY in Fig. \ref{fig:dGNY_RG} merges with the critical GNY model without a defect. Consequently, the violation occurs when either dWF or dG are deformed by the Yukawa coupling in the bulk because both DCFTs with negative anomaly flow towards the critical GNY model without an anomaly. This scenario is depicted in Fig. \ref{fig:dGNY_RG12}.

Finally, when $N>12$, the coupling $\ggn$ becomes negative and gradually approaches zero as $N$ tends to infinity. In this range, the $b$-anomaly of the dGNY model acquires small positive values. This time, the flow from the Gaussian fixed point (G) without a defect still terminates at the dGNY, thereby violating the naive generalization of the $b$-theorem. This case warrants caution, as a negative $\ggn$ implies the possibility of nontrivial classical profiles in the bulk \cite{Giombi2023,RavivMoshe2023}, accompanied by symmetry breaking. While studying this phenomenon is interesting, we have chosen to leave it outside the scope of the current work.
	
\section{Discussion}
\label{discussion}
		In this paper, we discussed a violation of the naive generalisation of the $b$-theorem, to the case of simultaneous bulk and defect RG flows, in a quantum system defined on a $d$-dimensional flat bulk coupled to a $2d$ defect. We demonstrated said violation in two distinct models namely, the $O(N)$ vector model and the Gross-Neveu-Yukawa (GNY) model, each coupled to a two-dimensional spherical defect. The computation of the $b$-anomaly at the various fixed points in each of these models was done using conformal perturbation theory, and the result was verified from an independent computation of the free energy in each system. 
	
	The failure to generalise the $b$-theorem to the case of simultaneous bulk-defect RG flows is reminiscent of a similar violation of the $g$-theorem \cite{Green:2007wr}, which characterises RG flows on $1d$ boundaries of $2d$ critical systems\footnote{The $g$-theorem has recently been shown to hold for $1d$ defects as well \cite{Cuomo:2021rkm}. It should be straightforward to show a violation of the defect $g$-theorem using arguments similar to \cite{Green:2007wr}.}. In case of $1d$ boundaries/defects, the $g$-function is interpreted as the \emph{impurity entropy}, that can either increase or decrease under bulk interactions. Such an interpretation provides a clear physical understanding underlying the failure to generalise the $g$-theorem to include bulk RG flows. Finding a similar understanding in the context of the $b$-theorem remains one of the interesting open problems.
	
	Another challenging open question that follows from such a violation, is regarding the existence and construction of an altogether different class of $C$- functions for defect QFTs, that follow from the partition function, and that again display monotonicity properties under simultaneous bulk and defect RG flows. 
	
	A completely independent, alternate construction of the $C$-functions\footnote{A proof for the $g$-theorem in $1+1$d CFTs using information-theoretic tools was first provided in \cite{Casini:2016fgb} and later generalised to boundaries in higher spacetime dimensions in \cite{Casini:2018nym}. The proof was finally extended to defects of arbitrary co-dimension in \cite{Casini:2023kyj}. For other works on defect RG flows using quantum information methods, see \cite{Yuan:2022oeo,Yuan:2023oni,Harper:2024aku}.} for boundaries and defects was provided using information theoretic techniques (such as relative entropy, entanglement entropy and the Quantum Null Energy Condition) in  \cite{Casini:2023kyj}. It will be interesting to explore whether such measures can capture the violation of the $g$-theorem and the $b$-theorem under simultaneous bulk-defect RG flows.\\
	
{\bf Acknowledgements}  We thank Chris Herzog and Maxime Tr\'{e}panier for helpful discussions and correspondence. TS and MS acknowledge partial support from the Israel Science Foundation (ISF) Center for Excellence grant (Grant Number 2289/18), BSF Grant No. 2022113, and NSF-BSF Grant No. 2022726. Additionally, MS acknowledges partial support from Israel's Council for Higher Education. RS is supported by the Royal Society-Newton International Fellowship NIF/R1/221054-Royal Society.
	
\appendix

\section{Computation of the $b$-anomaly}
\label{F-Appx}

	Another method of calculating the $b$-anomaly is based on evaluating the free energy. For a two-dimensional spherical conformal defect embedded in flat space, the free energy is given by,
	\begin{equation}
		\mathcal{F}=\mathcal{F}_\text{\tiny CFT}-\f{b}{3}\log\left(\mu R\right)+\ldots ~.
	\end{equation}
Here, $\mathcal{F}_\text{\tiny CFT}$ represents the free energy of the ambient CFT without the defect, $R$ corresponds to the radius of the sphere, and the ellipsis denotes non-universal terms that depend on the specific scheme used.
	
	In this appendix, we perform a direct evaluation of the free energy for the $O(N)$ vector model \eqref{ONaction} and the GNY model \eqref{GNYaction}, from which we extract the values of $b\text{\tiny{WF}}$ and $b_\text{\tiny{GNY}}$, respectively. The final results agree with \eqref{b_WF} and \eqref{b_GNY}, derived based on perturbative calculations of the defect stress tensor two-point function \cite{Shachar:2022fqk}.
	
Starting from the Gaussian $O(N)$ vector model \eqref{ONaction} without defects, we introduce an $O(N)$-invariant quartic interaction throughout the bulk and a $\phi^2$ term supported solely on the two-dimensional sphere of radius $R$. This deformation drives the system towards the $(\gwf,\lwf)$ fixed-point in the deep infrared (IR). As illustrated in Fig.\ref{Fig1}, this fixed-point is the only admissible IR endpoint of the RG flow when both deformations are present. Due to the weak coupling of the fixed point, we can expand the free energy in the IR as follows
	\begin{align} \label{freeE}
		\mathcal{F} &=- \log\left[\int\,[\md\phi]~ e^{-S_{CFT}-g_0\int_{\mathcal{S}^2}\phi^2 - \l_0\int_\mathcal{B}~(\phi^2)^2}\right] \nn\\
		&= \mathcal{F}_{\l_0}  -\log\left(1-g_0\int_{\mathcal{S}^2}\lan\phi^2\ran_{\l_0} +\f{g_0^2}{2}\int_{\mathcal{S}^2}\int_{\mathcal{S}^2}\lan\phi^2(\s_1)\phi^2(\s_2)\ran_{\l_0}\right.
		\nn\\
		&\left.-\f{g_0^3}{3!}\int_{\mathcal{S}^2}\int_{\mathcal{S}^2}\int_{\mathcal{S}^2}\lan\phi^2(\s_1)\phi^2(\s_2)\phi^2(\s_3)\ran_{\l_0}+\dots\right)
		\nn\\
		&= \mathcal{F}_{\l_0} - \f{g_0^2}2\int_{\mathcal{S}^2}\int_{\mathcal{S}^2}\lan\phi^2(\s_1)\phi^2(\s_2)\ran_0 + \f{g_0^2\l_0}{2}\int_{\mathcal{S}^2}\int_{\mathcal{S}^2}\int_\mathcal{B}~\lan\phi^2(\s_1)\phi^2(\s_2)(\phi^2)^2(x_3)\ran_0\nn\\
		&+\f{g_0^3}{3!}\int_{\mathcal{S}^2}\int_{\mathcal{S}^2}\int_{\mathcal{S}^2}\lan\phi^2(\s_1)\phi^2(\s_2)\phi^2(\s_3)\ran_0 + \ldots ~,
	\end{align}
where $\mathcal{F}_{\l_0} =\mathcal{F}|_{g_0=0}$ is anomaly-free, and in the last equality of \eqref{freeE}, we dropped $\lan\phi^2\ran_{\l_0}$ because this term is power-law divergent, and therefore it vanishes in dimensional regularization. Thus,	
	\begin{align} \label{Fexp}
		\Delta\mathcal{F}\equiv \mathcal{F} - \mathcal{F}_{\l_0}= -\f{g_0^2}2 J_1 + \f{g_0^2\l_0}{2}J_2 + \f{g_0^3}{3!}J_3 + \ldots ~.
	\end{align}
Here,
\begin{eqnarray}
 J_1 &=& \int_{\mathcal{S}^2}\int_{\mathcal{S}^2} \lan\phi^2(\s_1) \phi^2(\s_2)\ran_0 ~,
 \nonumber \\
 J_2 &=& \int_{\mathcal{S}^2}\int_{\mathcal{S}^2}\int_{\mathcal{B}} \lan\phi^2(\s_1) \phi^2(\s_2)\phi^3(x_3)\ran_0 ~,
 \\
 J_3 &=& \int_{\mathcal{S}^2}\int_{\mathcal{S}^2}\int_{\mathcal{S}^2} \lan\phi^2(\s_1) \phi^2(\s_2)\phi^3(\s_3)\ran_0 ~.
 \nonumber
\end{eqnarray}
Before proceeding with the computation of these expressions, let us review some useful identities. Denoting the length of the cord connecting two points $\sigma_1,\sigma_2$ on $\mathcal{S}^2$ embedded in $\mathbb{R}^d$ as $s\left(\sigma_1,\sigma_2\right)$, the following integrals hold \cite{Cardy:1988cwa,Shachar:2022fqk,Klebanov2011},
	\begin{align}
		&I_1\left(\alpha\right)=\int_{\mathcal{S}^2}\int_{\mathcal{S}^2}\f{1}{s\left(\sigma_1,\sigma_2\right)^{2\alpha}}
		=\frac{\pi ^2 \left(2R\right)^{4-2 \alpha }}{1-\alpha} ~, \\
		&I_2\left(\alpha\right) = \int_{\mathcal{S}^2}\int_{\mathcal{S}^2}\int_{\mathcal{S}^2}  \f{1}{\left[s(\sigma_1,\sigma_2)s(\sigma_2,\sigma_3)s(\sigma_3,\sigma_1)\right]^{\alpha}}=
		\frac{8\pi ^{9/2} \Gamma \left(2-\frac{3 \alpha
			}{2}\right) R^{6-3 \alpha }}{\Gamma
			\left(\frac{3-\alpha }{2}\right)^3} ~.
	\end{align}
In addition, we are going to use the conformal merging relation \cite{Goykhman2021},
	\begin{equation}
		\int d^d x_2 \f{1}{\left|x_{12} \right|^{2\Delta_1}\left|x_{23}\right|^{2\Delta_2}} =
		\f{U\left(\Delta_1,\Delta_2,d-\Delta_1-\Delta_2\right)}{\left|x_{13}\right|^{2\Delta_1+2\Delta_2-d}}~,
		\label{merging}
	\end{equation}
where $x_{12}=x_1-x_2$ and
	\begin{equation}
		U\left(\Delta_1,\Delta_2,\Delta_3\right)=\pi^\f{d}{2} \f{\Gamma\left(\f{d}{2}-\Delta_1\right)\Gamma\left(\f{d}{2}-\Delta_2\right)\Gamma\left(\f{d}{2}-\Delta_3\right)}{\Gamma\left(\Delta_1\right)\Gamma\left(\Delta_2\right)\Gamma\left(\Delta_3\right)} ~.
	\end{equation}
Using the above identities, we obtain 
\begin{eqnarray}
J_1 &=&2NC_\phi^2~I_1\left(2-\eps\right) = - (2R)^{2\eps} \f{N}{8\pi^2} + \ldots ~,
\nonumber \\
J_2 &=&  8N(N+2)C_\phi^4 \, U\left(2-\eps,2-\eps,\eps\right) I_1\Big(2-\f{3}2\eps\Big)=-(2R)^{3\eps} \f{N(N+2)}{8\pi^4\eps}+ \ldots ~,
\nonumber \\
J_3&=& 8NC_{\phi}^3~I_2(2-\eps)=-R^{3\eps} \left(\f{2N}{3\pi^3\eps}\right) + \ldots ~.
\end{eqnarray}
Combining everything and using \eqref{bareVSren} and \eqref{freeOPE} to express the bare parameters in \eqref{Fexp} in terms of the renormalized ones yields,\footnote{This expression is not finite in the limit $\epsilon \to 0$ because we did not include in $\mathcal{F}$ the contribution of the geometric counterterm proportional to the integral of the Ricci scalar over the defect.}
	\begin{align}
		\Delta\mathcal{F} =& \f{N}{8\pi^2}\f{g^2}2\left(2\mu R\right)^{2\eps}\left(1+\f{g}{\pi\eps}  +
		\f{\l(N+2)}{2\pi^2\eps}\right)^2  \nn\\
		& -
		(2\mu R)^{3\eps} \f{g^2\l}{2}\f{N(N+2)}{8\pi^4\eps} - (\mu R)^{3\eps}\f{g^3}{6}\left(\f{2N}{3\pi^3\eps}\right) + \ldots ~.
	\end{align}
Identifying the coefficient of $\log(\mu R)$ and plugging in the critical values \eqref{criticalWF} yields,
	\begin{align}
		b_\text{\tiny{WF}}&= \left(-\f{3N\eps}{8\pi^2}g^2  +\f{N}{4\pi^3}g^3 +\f{3N(N+2)}{16\pi^4}g^2\l\right)\bigg|_{\l_* = \lwf,~ g_*=\gwf}\nn\\
		&=-\f{27 N}{(N+8)^3}\eps^3 +\ldots ~.
	\end{align}
	In agreement with \eqref{b_WF}.
	
To conclude this appendix, we compute the free energy for the GNY model. The computation follows in the same way as before, with an additional integral that requires evaluation,
	\begin{align} \label{FGNY}
		\Delta\mathcal{F} = -\f{g_0^2}2 J_1 + \f{g_0^2\l_0}{2 \cdot 4!}J_2 + \f{g_0^3}{3!}J_3
		-\f{g_0^2\yuk_0^2}{4}J_4 +\ldots ~.
	\end{align}
This integral is given by
\begin{align}
		J_4= &\int_{\mathcal{S}^2}\int_{\mathcal{S}^2}\int_\mathcal{B}\int_\mathcal{B}~\lan\phi_0^2\left(\s_1\right)\phi_0^2\left(\s_2\right)\bar{\psi}_0\psi_0\phi_0\left(x_1\right)\bar{\psi}_0\psi_0\phi_0\left(x_2\right)\ran_0 \nn \\
		&=8 N \, C^3_\phi C^2_\psi~U\left(\f{2-\eps}{2},3-\eps,\f{\eps}{2}\right) U\left(\f{2-\eps}{2},2-\eps,1+\f{\eps}{2}\right)I_1\left(2-\f{3\eps}{2}\right) \nn \\
		&= (2R)^{3\eps}\f{N}{32\pi^4 \eps} + \ldots ~,
	\end{align}
where we employed the conformal merging relation \eqref{merging} twice, and used the position space representation of the two-point function for a massless Dirac field,
\begin{equation}
 \langle \psi(x_1) \bar\psi(x_2) \rangle =C_\psi {\slashed{x}_{12} \over |x_{12}|^{d}} ~, \quad C_\psi={\Gamma\({d\over 2}\)\over 2\pi^{d\over 2}}~.
 \label{Dirac2p}
\end{equation}	
In addition, the Yukawa interaction contributes to the first term of \eqref{FGNY} through the following renormalization of the coupling (see Appendix \ref{GNY-RG})
	\begin{equation}
		g_0=g\mu^\eps \(1+\f{g}{\pi\eps}+\f{\l}{16\pi^2\eps}+\f{N\yuk^2}{16\pi^2\eps}+ \ldots \)~.
	\end{equation}
Overall, we find,
	\begin{align}
		b_\text{\tiny{GNY}}&= \left(-\f{3\eps}{8\pi^2}g^2  +\f{1}{4\pi^3}g^3 + \f{3}{128\pi^4}g^2 \l
		+ \f{3Ng^2\yuk^2}{128\pi^4}\right)\bigg|_{\yuk_*=\ygn,~\l_* = \lgn,~ g_*=\ggn} + \ldots 
		\nn\\
		&=- { \epsilon^3\over 8} \bigg(1-\f{6+5N+\sqrt{N^2+132N+36}}{6\left(N+6\right) } \,\bigg)^3 +\ldots ~.
	\end{align}
This expression agrees with \eqref{b_GNY}.

\section{RG beta functions for GNY model with defect}
\label{GNY-RG}

To renormalize the GNY model, we supplement its action (\ref{GNYaction}) with the following counterterms
\begin{equation}
 S_\text{c.t.}=\int_\mathcal{B} ~ \Big( \delta Z_\psi\bar{\psi}\slashed{\partial}\psi +\f{\delta Z_\phi}{2}(\partial\phi)^2 +\delta Z_\yuk \, \yuk\, \mu^\f{\eps}{2} \, \bar{\psi}\psi\phi
 +\f{\delta Z_\l}{4!} \, \l  \, \mu^\eps  \phi^4\Big) + \delta Z_g \, g \,\mu^\eps \int_{\mathcal{S}^2}\phi^2 ~,
\end{equation}
where all couplings are dimensionless and $\delta Z_i$ for $i=\psi, \phi, \yuk, \l, g$ contain an ascending series of poles in $1/\epsilon$; that is, we employ the minimal subtraction scheme. Define $Z_i=1+\delta Z_i$, then the relations between the bare and renormalized parameters are given by
\begin{gather}
		\psi_0=\sqrt{Z_\psi} \, \psi,\qquad\phi_0=\sqrt{Z_\phi} \, \phi  ~,\\
		\yuk_0=  {Z_\yuk\over \sqrt{Z_\phi} Z_\psi } \, \mu^\f{\eps}{2} \, \yuk ,\qquad \l_0= {Z_\l\over Z^2_\phi} ~ \mu^\eps  \l,
		\qquad g_0= {Z_g \over Z_\phi} ~ \mu^\eps g ~.
		\label{CTdefs}
\end{gather}
In this appendix, we carry out explicit calculations of the 1-loop Feynman graphs contributing to renormalization constants $Z_i$.

\begin{figure}[t!]
		\centering
		\includegraphics[width=0.7\columnwidth]{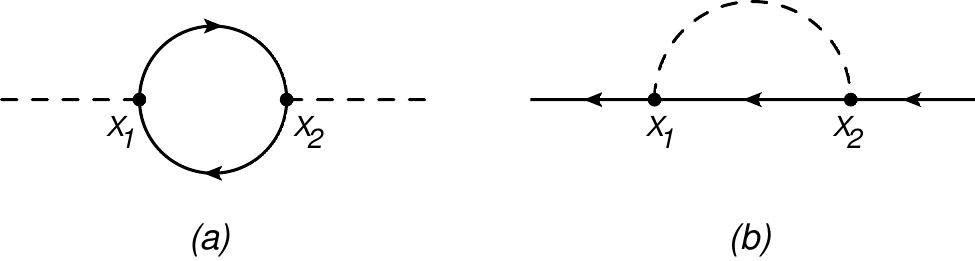}
		\caption{One-loop diagrams contributing to the wave function renormalization in the GNY model for \textbf{(a)} the scalar field $\phi$ and \textbf{(b)} the Dirac field $\psi$. Internal solid lines with arrows correspond to the propagator of the Dirac field, while internal dashed lines represent the propagator of the scalar field. External dashed and solid lines denote scalar and Dirac background fields, respectively.}
		\label{fig:wave}
	\end{figure}

The wave function renormalization of the scalar and Dirac fields are associated with the diagrams in Fig.\ref{fig:wave},
\begin{eqnarray}
 \text{Fig.}\ref{fig:wave}\text{(a)}&=& - {N_\text{\tiny F} \yuk^2 \mu^{\eps} \, C_\psi^2 \over 2} \int_\mathcal{B} \int_\mathcal{B} \phi(x_1) \phi(x_2)\text{tr} \Big( {\slashed{x}_{12}  \slashed{x}_{21}\over |x_{12}|^{d} |x_{21}|^d}\Big) = - {N\yuk^2 \over 32\pi^2 \eps} \int_\mathcal{B} \del\phi \del\phi + \ldots ~,
 \nonumber
  \\
 \text{Fig.}\ref{fig:wave}\text{(b)}&=& \yuk^2 \mu^\eps  C_\psi C_\phi   \int_\mathcal{B}  \int_\mathcal{B} {\bar\psi(x_1) \slashed{x}_{12} \psi(x_2) \over (x_{12}^2)^{d-1}}
 =-{\yuk^2 \over 16\pi^2 \eps} \int_\mathcal{B}\bar{\psi}\slashed{\partial}\psi + \ldots ~.
 \nonumber
\end{eqnarray}
where $x_{12}=x_1-x_2$, $N=4N_\text{F}$ is the total number of fermionic degrees of freedom in four dimensions, the trace is done over the spinor indices, and we used \eqref{Dirac2p}. Hence,
\begin{eqnarray} 
	&&Z_\phi=1- \f{N\yuk^2}{16\pi^2\eps}+ \ldots  ~,\label{zdef} 	\nonumber \\
	&& Z_\psi= 1- {\eta^2\over 16\pi^2 \eps} + \ldots ~.
	\label{wave-renorm} 
\end{eqnarray}

\begin{figure}[t!]
		\centering
		\includegraphics[width=0.4\columnwidth]{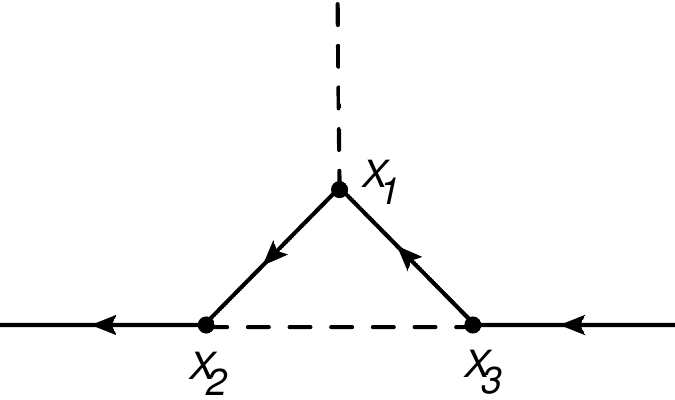}
		\caption{One-loop diagram contributing to the renormalization of the Yukawa coupling $\eta$ in the GNY model. Internal solid lines with arrows correspond to the propagator of the Dirac field $\psi$, while the internal dashed line represents the propagator of the scalar field $\phi$. In contrast, external dashed and solid lines denote scalar and Dirac background fields, respectively.}
		\label{fig:Zeta}
	\end{figure}

Next, we fix $Z_\yuk$ by evaluating the one-loop diagram in Fig. \ref{fig:Zeta},
\begin{equation}
  \text{Fig.}\ref{fig:Zeta}= - \yuk^3 \mu^{3\epsilon\over 2} C_\psi^2 C_\phi \int_\mathcal{B} \int_\mathcal{B}  \int_\mathcal{B} \phi(x_1)
  {\bar\psi(x_2)\slashed{x}_{21} \slashed{x}_{13} \psi(x_3)\over |x_{21}|^d|x_{31}|^d|x_{23}|^{d-2}} ~.
\end{equation} 
The pole in $\epsilon$ of this integral is obtained by expanding the Dirac fields around $x_1$ and subsequently integrating over $x_2$ and $x_3$ using the scalar-fermion propagator merging relation of the form
\begin{eqnarray}
 \int d^d x_3 {\slashed{x}_{31} \over |x_{31}|^{2\Delta_1+1} |x_{23}|^{2\Delta_2}} &=&
 { \pi^{d\over 2}  \Gamma\( {d\over 2} -  \Delta_2\)  \Gamma\({d\over 2}-\Delta_1+{1\over 2}\) \Gamma\(\Delta_1+\Delta_2-{d\over 2}+{1\over 2}\) \over \Gamma(\Delta_2) \Gamma\(\Delta_1+{1\over 2}\) \Gamma\(d-\Delta_1-\Delta_2+{1\over 2}\)} 
 \nonumber \\
 &\times&{\slashed{x}_{21} \over |x_{12}|^{2(\Delta_1+\Delta_2)-d+1}} ~.
 \label{mergingFS}
\end{eqnarray}
The final expression is given by
\begin{equation}
  \text{Fig.}\ref{fig:Zeta}=   {\yuk^3 \mu^{\eps\over 2}\over 8\pi^2\epsilon} \int_\mathcal{B} \bar\psi\psi\phi + \ldots ~.
\end{equation} 
Thus
\begin{equation}
 Z_\yuk=1 + {\yuk^2 \over 8\pi^2\epsilon} + \ldots ~.
 \label{eta-renorm}
\end{equation}
Substituting (\ref{wave-renorm}) and (\ref{eta-renorm}) into the relation between $\eta_0$ and $\eta$ in (\ref{CTdefs}), gives \cite{ZinnJustin1991,Moshe2003}
\begin{equation}
 \yuk_0=  {Z_\yuk\over \sqrt{Z_\phi} Z_\psi } \, \mu^\f{\eps}{2} \, \yuk \quad \Rightarrow \quad \beta_{\yuk^2} = -\eps \yuk^2+\f{N+6}{16\pi^2}\yuk^4 ~.
\end{equation}

\begin{figure}[t!]
		\centering
		\includegraphics[width=0.6\columnwidth]{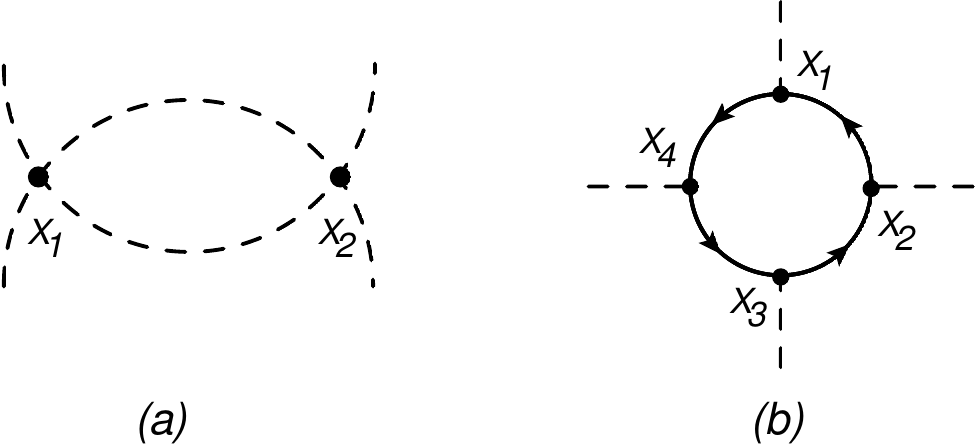}
		\caption{Two one-loop diagrams contributing to the renormalization of the coupling $\lambda$ in the GNY model. Internal solid lines with arrows correspond to the propagator of the Dirac field $\psi$, while the internal dashed lines represent the propagator of the scalar field $\phi$. External dashed and solid lines denote scalar and Dirac background fields, respectively.}
		\label{fig:Zlam}
	\end{figure}

To derive the RG flow for $\l$, one has to consider the one-loop diagrams shown in Fig. \ref{fig:Zlam}
\begin{eqnarray}
 \text{Fig.}\ref{fig:Zlam}\text{(a)}&=& {\l^2\mu^{2\eps} C_\phi^2\over 16}\int_\mathcal{B}\int_\mathcal{B} {\phi^2(x_1) \phi^2(x_2)\over |x_{12}|^{2(d-2)}}  ~,
  \\
 \text{Fig.}\ref{fig:Zlam}\text{(b)}&=& - {\yuk^4 \mu^{2\epsilon} C_\psi^4 N_\text{\tiny F}\over 4}  \int_\mathcal{B} \int_\mathcal{B} \int_\mathcal{B} \int_\mathcal{B} \prod_{i=1}^4 \phi(x_i)
  \text{tr} \( {\slashed{x}_{12} \slashed{x}_{23} \slashed{x}_{34} \slashed{x}_{41} \over |x_{12}|^d|x_{23}|^d|x_{34}|^d|x_{41}|^d} \)  ~.
   \nonumber
\end{eqnarray}
The counterterm $Z_\l$ is entirely fixed by the poles in $\epsilon$, therefore we expand the background fields around $x_1$ and get
\begin{eqnarray}
 \text{Fig.}\ref{fig:Zlam}\text{(a)}&=&  {\l^2\mu^{\eps} \over 128 \pi^2 \epsilon}\int_\mathcal{B} \phi^4 +\ldots  ~,
 \nonumber
  \\
 \text{Fig.}\ref{fig:Zlam}\text{(b)}&=& -{\eta^4 \mu^\eps N\over 32 \pi^2 \eps} \int_\mathcal{B} \phi^4  +\ldots  ~.
\end{eqnarray}
To get the above expression for \text{Fig.} \ref{fig:Zlam}\text{(b)}, we used (\ref{mergingFS}) and the merging relation for the pair of Dirac propagators,
\begin{eqnarray}
 \int d^d x_3 {\slashed{x}_{13}\slashed{x}_{32} \over |x_{32}|^{2\Delta_1+1} |x_{13}|^{2\Delta_2+1}} &=&
 {- \pi^{d\over 2}  \Gamma\( \Delta_1 + \Delta_2-{d\over 2}\)  \Gamma\({d\over 2}-\Delta_1+{1\over 2}\) \Gamma\({d\over 2}-\Delta_2+{1\over 2}\) \over \Gamma(d-\Delta_1-\Delta_2) \Gamma\(\Delta_1+{1\over 2}\) \Gamma\(\Delta_2+{1\over 2}\)} 
 \nonumber \\
 &\times&{\mathbb{I}_{[{d\over2}]\times [{d\over2}]} \over |x_{12}|^{2(\Delta_1+\Delta_2)-d}} ~,
 \label{mergingFF}
\end{eqnarray}
where $\mathbb{I}$ is an identity matrix in the spinor space. Hence,
\begin{equation}
 Z_\l=1 + {3\l \over 16 \pi^2 \eps} - {3 \eta^4 N\over 4 \pi^2 \lambda \, \eps} + \ldots ~.
\end{equation}
As a result, we obtain the following RG flow for $\l$ \cite{ZinnJustin1991,Moshe2003},
	\begin{gather}
		\l_0= {Z_\l\over Z^2_\phi} ~ \mu^\eps  \l \quad \Rightarrow \quad \beta_\l = -\eps\l +\f{1}{8\pi^2}\Big(\f{3}{2}\l^2+N\l \yuk^2-6N\yuk^4\Big) ~.
	\end{gather}

\begin{figure}[t!]
		\centering
		\includegraphics[width=0.6\columnwidth]{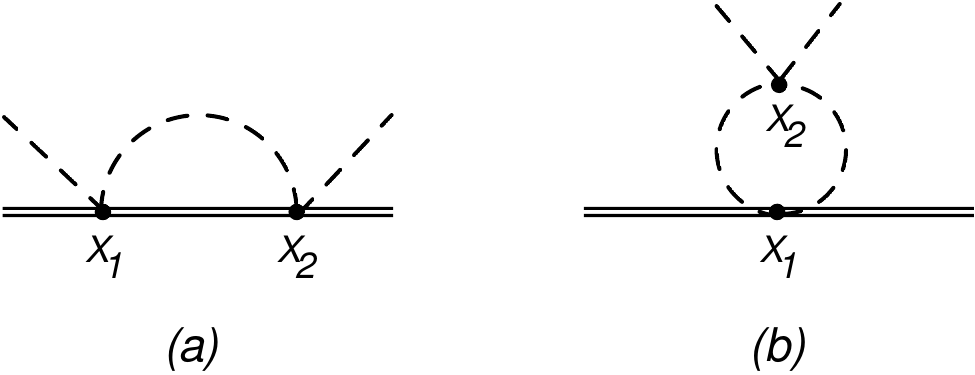}
		\caption{Two one-loop diagrams contributing to the renormalization of the defect coupling $g$ in the GNY model. The double line represents the two-dimensional defect, while the internal and external dashed lines represent the propagator of the scalar field $\phi$ and external scalar background, respectively.}
		\label{fig:Zg}
	\end{figure}
	
Finally, the leading non-trivial corrections to $Z_g$ come from the two diagrams shown in Fig. \ref{fig:Zg},
\begin{eqnarray}
 \text{Fig.}\ref{fig:Zg}\text{(a)}&=& 2\, C_\phi\,\mu^{2\epsilon} g^2 \int_{\mathcal{S}^2} \int_{\mathcal{S}^2} {\phi(x_1) \phi(x_2)\over |x_1-x_2|^{2-\epsilon}}
 =  {\mu^{\epsilon} g^2 \over \pi \epsilon} \int_{\mathcal{S}^2} \phi^2(x_1)  + \ldots
  \\
 \text{Fig.}\ref{fig:Zg}\text{(b)}&=& {C_\phi^2 \, \mu^{2\epsilon} \lambda g\over 2} \int_{\mathcal{S}^2} \int d^dx_2 \, {\phi^2(x_2)\over |x_1-x_2|^{4-2\epsilon}} = {\mu^{\epsilon} g \lambda \over 16 \pi^2 \epsilon} \int_{\mathcal{S}^2} \phi^2(x_1)  + \ldots ~.
 \nonumber
\end{eqnarray}
Hence,
\begin{equation}
 Z_g=1+\f{g}{\pi\eps}+\f{\l}{16\pi^2\eps} + \ldots ~.
 \label{g-renorm}
\end{equation}
Substituting $Z_g$ and $Z_\phi$ from (\ref{wave-renorm}) into the relation between the bare $g_0$ and renormalized $g$ in (\ref{CTdefs}) yields
\begin{equation}
 g_0= {Z_g \over Z_\phi} ~ \mu^\eps g \quad \Rightarrow \quad \beta_g = -\eps g + \f{g^2}{\pi} + \f{g}{16\pi^2}\left(\l+N\yuk^2\right) ~.
\end{equation}
This beta function is new; it was not considered before in the literature.

\bibliographystyle{utphys.bst}
\bibliography{ref.bib}

\providecommand{\href}[2]{#2}\begingroup\raggedright\begin{thebibliography}{10}

\bibitem{Zamolodchikov:1986gt}
A.~B. Zamolodchikov, ``{Irreversibility of the Flux of the Renormalization
  Group in a 2D Field Theory},'' {\em JETP Lett.} {\bfseries 43} (1986)
  730--732.

\bibitem{Cardy:1988cwa}
J.~L. Cardy, ``{Is There a c Theorem in Four-Dimensions?},''
  \href{http://dx.doi.org/10.1016/0370-2693(88)90054-8}{{\em Phys. Lett. B}
  {\bfseries 215} (1988) 749--752}.

\bibitem{Cappelli:1990yc}
A.~Cappelli, D.~Friedan, and J.~I. Latorre, ``{C theorem and spectral
  representation},'' \href{http://dx.doi.org/10.1016/0550-3213(91)90102-4}{{\em
  Nucl. Phys. B} {\bfseries 352} (1991) 616--670}.

\bibitem{Osborn:1991gm}
H.~Osborn, ``{Weyl consistency conditions and a local renormalization group
  equation for general renormalizable field theories},''
  \href{http://dx.doi.org/10.1016/0550-3213(91)80030-P}{{\em Nucl. Phys. B}
  {\bfseries 363} (1991) 486--526}.

\bibitem{Myers:2010tj}
R.~C. Myers and A.~Sinha, ``{Holographic c-theorems in arbitrary dimensions},''
  \href{http://dx.doi.org/10.1007/JHEP01(2011)125}{{\em JHEP} {\bfseries 01}
  (2011) 125}, \href{http://arxiv.org/abs/1011.5819}{{\ttfamily arXiv:1011.5819
  [hep-th]}}.

\bibitem{Jafferis:2011zi}
D.~L. Jafferis, I.~R. Klebanov, S.~S. Pufu, and B.~R. Safdi, ``{Towards the
  F-Theorem: N=2 Field Theories on the Three-Sphere},''
  \href{http://dx.doi.org/10.1007/JHEP06(2011)102}{{\em JHEP} {\bfseries 06}
  (2011) 102}, \href{http://arxiv.org/abs/1103.1181}{{\ttfamily arXiv:1103.1181
  [hep-th]}}.

\bibitem{Komargodski:2011vj}
Z.~Komargodski and A.~Schwimmer, ``{On Renormalization Group Flows in Four
  Dimensions},'' \href{http://dx.doi.org/10.1007/JHEP12(2011)099}{{\em JHEP}
  {\bfseries 12} (2011) 099}, \href{http://arxiv.org/abs/1107.3987}{{\ttfamily
  arXiv:1107.3987 [hep-th]}}.

\bibitem{Komargodski:2011xv}
Z.~Komargodski, ``{The Constraints of Conformal Symmetry on RG Flows},''
  \href{http://dx.doi.org/10.1007/JHEP07(2012)069}{{\em JHEP} {\bfseries 07}
  (2012) 069}, \href{http://arxiv.org/abs/1112.4538}{{\ttfamily arXiv:1112.4538
  [hep-th]}}.

\bibitem{Casini:2012ei}
H.~Casini and M.~Huerta, ``{On the RG running of the entanglement entropy of a
  circle},'' \href{http://dx.doi.org/10.1103/PhysRevD.85.125016}{{\em Phys.
  Rev. D} {\bfseries 85} (2012) 125016},
  \href{http://arxiv.org/abs/1202.5650}{{\ttfamily arXiv:1202.5650 [hep-th]}}.

\bibitem{Elvang:2012st}
H.~Elvang, D.~Z. Freedman, L.-Y. Hung, M.~Kiermaier, R.~C. Myers, and
  S.~Theisen, ``{On renormalization group flows and the a-theorem in 6d},''
  \href{http://dx.doi.org/10.1007/JHEP10(2012)011}{{\em JHEP} {\bfseries 10}
  (2012) 011}, \href{http://arxiv.org/abs/1205.3994}{{\ttfamily arXiv:1205.3994
  [hep-th]}}.

\bibitem{Yonekura:2012kb}
K.~Yonekura, ``{Perturbative c-theorem in d-dimensions},''
  \href{http://dx.doi.org/10.1007/JHEP04(2013)011}{{\em JHEP} {\bfseries 04}
  (2013) 011}, \href{http://arxiv.org/abs/1212.3028}{{\ttfamily arXiv:1212.3028
  [hep-th]}}.

\bibitem{Grinstein:2013cka}
B.~Grinstein, A.~Stergiou, and D.~Stone, ``{Consequences of Weyl Consistency
  Conditions},'' \href{http://dx.doi.org/10.1007/JHEP11(2013)195}{{\em JHEP}
  {\bfseries 11} (2013) 195}, \href{http://arxiv.org/abs/1308.1096}{{\ttfamily
  arXiv:1308.1096 [hep-th]}}.

\bibitem{Jack:2013sha}
I.~Jack and H.~Osborn, ``{Constraints on RG Flow for Four Dimensional Quantum
  Field Theories},''
  \href{http://dx.doi.org/10.1016/j.nuclphysb.2014.03.018}{{\em Nucl. Phys. B}
  {\bfseries 883} (2014) 425--500},
  \href{http://arxiv.org/abs/1312.0428}{{\ttfamily arXiv:1312.0428 [hep-th]}}.

\bibitem{Giombi:2014xxa}
S.~Giombi and I.~R. Klebanov, ``{Interpolating between $a$ and $F$},''
  \href{http://dx.doi.org/10.1007/JHEP03(2015)117}{{\em JHEP} {\bfseries 03}
  (2015) 117}, \href{http://arxiv.org/abs/1409.1937}{{\ttfamily arXiv:1409.1937
  [hep-th]}}.

\bibitem{Jack:2015tka}
I.~Jack, D.~R.~T. Jones, and C.~Poole, ``{Gradient flows in three
  dimensions},'' \href{http://dx.doi.org/10.1007/JHEP09(2015)061}{{\em JHEP}
  {\bfseries 09} (2015) 061}, \href{http://arxiv.org/abs/1505.05400}{{\ttfamily
  arXiv:1505.05400 [hep-th]}}.

\bibitem{Cordova:2015fha}
C.~Cordova, T.~T. Dumitrescu, and K.~Intriligator, ``{Anomalies,
  renormalization group flows, and the a-theorem in six-dimensional (1, 0)
  theories},'' \href{http://dx.doi.org/10.1007/JHEP10(2016)080}{{\em JHEP}
  {\bfseries 10} (2016) 080}, \href{http://arxiv.org/abs/1506.03807}{{\ttfamily
  arXiv:1506.03807 [hep-th]}}.

\bibitem{Casini:2015woa}
H.~Casini, M.~Huerta, R.~C. Myers, and A.~Yale, ``{Mutual information and the
  F-theorem},'' \href{http://dx.doi.org/10.1007/JHEP10(2015)003}{{\em JHEP}
  {\bfseries 10} (2015) 003}, \href{http://arxiv.org/abs/1506.06195}{{\ttfamily
  arXiv:1506.06195 [hep-th]}}.

\bibitem{Casini:2017vbe}
H.~Casini, E.~Test\'e, and G.~Torroba, ``{Markov Property of the Conformal
  Field Theory Vacuum and the a Theorem},''
  \href{http://dx.doi.org/10.1103/PhysRevLett.118.261602}{{\em Phys. Rev.
  Lett.} {\bfseries 118} no.~26, (2017) 261602},
  \href{http://arxiv.org/abs/1704.01870}{{\ttfamily arXiv:1704.01870
  [hep-th]}}.

\bibitem{Fluder:2020pym}
M.~Fluder and C.~F. Uhlemann, ``{Evidence for a 5d F-theorem},''
  \href{http://dx.doi.org/10.1007/JHEP02(2021)192}{{\em JHEP} {\bfseries 02}
  (2021) 192}, \href{http://arxiv.org/abs/2011.00006}{{\ttfamily
  arXiv:2011.00006 [hep-th]}}.

\bibitem{Delacretaz:2021ufg}
L.~V. Delacretaz, A.~L. Fitzpatrick, E.~Katz, and M.~T. Walters,
  ``{Thermalization and hydrodynamics of two-dimensional quantum field
  theories},'' \href{http://dx.doi.org/10.21468/SciPostPhys.12.4.119}{{\em
  SciPost Phys.} {\bfseries 12} no.~4, (2022) 119},
  \href{http://arxiv.org/abs/2105.02229}{{\ttfamily arXiv:2105.02229
  [hep-th]}}.

\bibitem{Cardy:1984bb}
J.~L. Cardy, ``{Conformal Invariance and Surface Critical Behavior},''
  \href{http://dx.doi.org/10.1016/0550-3213(84)90241-4}{{\em Nucl. Phys. B}
  {\bfseries 240} (1984) 514--532}.

\bibitem{Cardy:1989ir}
J.~L. Cardy, ``{Boundary Conditions, Fusion Rules and the Verlinde Formula},''
  \href{http://dx.doi.org/10.1016/0550-3213(89)90521-X}{{\em Nucl. Phys. B}
  {\bfseries 324} (1989) 581--596}.

\bibitem{McAvity:1995zd}
D.~M. McAvity and H.~Osborn, ``{Conformal field theories near a boundary in
  general dimensions},''
  \href{http://dx.doi.org/10.1016/0550-3213(95)00476-9}{{\em Nucl. Phys. B}
  {\bfseries 455} (1995) 522--576},
  \href{http://arxiv.org/abs/cond-mat/9505127}{{\ttfamily
  arXiv:cond-mat/9505127}}.

\bibitem{Affleck:1995ge}
I.~Affleck, ``{Conformal field theory approach to the Kondo effect},'' {\em
  Acta Phys. Polon. B} {\bfseries 26} (1995) 1869--1932,
  \href{http://arxiv.org/abs/cond-mat/9512099}{{\ttfamily
  arXiv:cond-mat/9512099}}.

\bibitem{Sachdev99}
S.~Sachdev, C.~Buragohain, and M.~Vojta, ``Quantum impurity in a nearly
  critical two dimensional antiferromagnet,''.
  \url{https://arxiv.org/abs/cond-mat/0004156}.

\bibitem{Vojta_2000}
M.~Vojta, C.~Buragohain, and S.~Sachdev, ``Quantum impurity dynamics in
  two-dimensional antiferromagnets and superconductors,''
  \href{http://dx.doi.org/10.1103/physrevb.61.15152}{{\em Physical Review B}
  {\bfseries 61} no.~22, (Jun, 2000) 15152--15184}.
  \url{https://doi.org/10.1103%2Fphysrevb.61.15152}.

\bibitem{Polchinski:2011im}
J.~Polchinski and J.~Sully, ``{Wilson Loop Renormalization Group Flows},''
  \href{http://dx.doi.org/10.1007/JHEP10(2011)059}{{\em JHEP} {\bfseries 10}
  (2011) 059}, \href{http://arxiv.org/abs/1104.5077}{{\ttfamily arXiv:1104.5077
  [hep-th]}}.

\bibitem{Gaiotto:2013nva}
D.~Gaiotto, D.~Mazac, and M.~F. Paulos, ``{Bootstrapping the 3d Ising twist
  defect},'' \href{http://dx.doi.org/10.1007/JHEP03(2014)100}{{\em JHEP}
  {\bfseries 03} (2014) 100}, \href{http://arxiv.org/abs/1310.5078}{{\ttfamily
  arXiv:1310.5078 [hep-th]}}.

\bibitem{Billo:2016cpy}
M.~Bill\`o, V.~Goncalves, E.~Lauria, and M.~Meineri, ``{Defects in conformal
  field theory},'' \href{http://dx.doi.org/10.1007/JHEP04(2016)091}{{\em JHEP}
  {\bfseries 04} (2016) 091}, \href{http://arxiv.org/abs/1601.02883}{{\ttfamily
  arXiv:1601.02883 [hep-th]}}.

\bibitem{Bianchi:2015liz}
L.~Bianchi, M.~Meineri, R.~C. Myers, and M.~Smolkin, ``{R\'enyi entropy and
  conformal defects},'' \href{http://dx.doi.org/10.1007/JHEP07(2016)076}{{\em
  JHEP} {\bfseries 07} (2016) 076},
  \href{http://arxiv.org/abs/1511.06713}{{\ttfamily arXiv:1511.06713
  [hep-th]}}.

\bibitem{Solodukhin:2015eca}
S.~N. Solodukhin, ``{Boundary terms of conformal anomaly},''
  \href{http://dx.doi.org/10.1016/j.physletb.2015.11.036}{{\em Phys. Lett. B}
  {\bfseries 752} (2016) 131--134},
  \href{http://arxiv.org/abs/1510.04566}{{\ttfamily arXiv:1510.04566
  [hep-th]}}.

\bibitem{Fursaev:2016inw}
D.~V. Fursaev and S.~N. Solodukhin, ``{Anomalies, entropy and boundaries},''
  \href{http://dx.doi.org/10.1103/PhysRevD.93.084021}{{\em Phys. Rev. D}
  {\bfseries 93} no.~8, (2016) 084021},
  \href{http://arxiv.org/abs/1601.06418}{{\ttfamily arXiv:1601.06418
  [hep-th]}}.

\bibitem{Lauria:2018klo}
E.~Lauria, M.~Meineri, and E.~Trevisani, ``{Spinning operators and defects in
  conformal field theory},''
  \href{http://dx.doi.org/10.1007/JHEP08(2019)066}{{\em JHEP} {\bfseries 08}
  (2019) 066}, \href{http://arxiv.org/abs/1807.02522}{{\ttfamily
  arXiv:1807.02522 [hep-th]}}.

\bibitem{Gadde:2016fbj}
A.~Gadde, ``{Conformal constraints on defects},''
  \href{http://dx.doi.org/10.1007/JHEP01(2020)038}{{\em JHEP} {\bfseries 01}
  (2020) 038}, \href{http://arxiv.org/abs/1602.06354}{{\ttfamily
  arXiv:1602.06354 [hep-th]}}.

\bibitem{Herzog:2020bqw}
C.~P. Herzog and A.~Shrestha, ``{Two point functions in defect CFTs},''
  \href{http://dx.doi.org/10.1007/JHEP04(2021)226}{{\em JHEP} {\bfseries 04}
  (2021) 226}, \href{http://arxiv.org/abs/2010.04995}{{\ttfamily
  arXiv:2010.04995 [hep-th]}}.

\bibitem{Giombi:2021uae}
S.~Giombi, E.~Helfenberger, Z.~Ji, and H.~Khanchandani, ``{Monodromy defects
  from hyperbolic space},''
  \href{http://dx.doi.org/10.1007/JHEP02(2022)041}{{\em JHEP} {\bfseries 02}
  (2022) 041}, \href{http://arxiv.org/abs/2102.11815}{{\ttfamily
  arXiv:2102.11815 [hep-th]}}.

\bibitem{Liu_2021}
S.~Liu, H.~Shapourian, A.~Vishwanath, and M.~A. Metlitski, ``Magnetic
  impurities at quantum critical points: large-$n$ expansion and spt
  connections,'' \href{http://dx.doi.org/10.1103/physrevb.104.104201}{{\em
  Physical Review B} {\bfseries 104} no.~10, (Sep, 2021) }.
  \url{https://doi.org/10.1103%2Fphysrevb.104.104201}.

\bibitem{Herzog:2022jqv}
C.~P. Herzog and A.~Shrestha, ``{Conformal surface defects in Maxwell theory
  are trivial},'' \href{http://dx.doi.org/10.1007/JHEP08(2022)282}{{\em JHEP}
  {\bfseries 08} (2022) 282}, \href{http://arxiv.org/abs/2202.09180}{{\ttfamily
  arXiv:2202.09180 [hep-th]}}.

\bibitem{Herzog:2019rke}
C.~P. Herzog and I.~Shamir, ``{How a-type anomalies can depend on marginal
  couplings},'' \href{http://dx.doi.org/10.1103/PhysRevLett.124.011601}{{\em
  Phys. Rev. Lett.} {\bfseries 124} no.~1, (2020) 011601},
  \href{http://arxiv.org/abs/1907.04952}{{\ttfamily arXiv:1907.04952
  [hep-th]}}.

\bibitem{Bianchi:2019umv}
L.~Bianchi, ``{Marginal deformations and defect anomalies},''
  \href{http://dx.doi.org/10.1103/PhysRevD.100.126018}{{\em Phys. Rev. D}
  {\bfseries 100} no.~12, (2019) 126018},
  \href{http://arxiv.org/abs/1907.06193}{{\ttfamily arXiv:1907.06193
  [hep-th]}}.

\bibitem{Herzog:2021hri}
C.~P. Herzog and I.~Shamir, ``{Anomalies from correlation functions in defect
  conformal field theory},''
  \href{http://dx.doi.org/10.1007/JHEP07(2021)091}{{\em JHEP} {\bfseries 07}
  (2021) 091}, \href{http://arxiv.org/abs/2103.06311}{{\ttfamily
  arXiv:2103.06311 [hep-th]}}.

\bibitem{Affleck:1991tk}
I.~Affleck and A.~W.~W. Ludwig, ``{Universal noninteger 'ground state
  degeneracy' in critical quantum systems},''
  \href{http://dx.doi.org/10.1103/PhysRevLett.67.161}{{\em Phys. Rev. Lett.}
  {\bfseries 67} (1991) 161--164}.

\bibitem{Yamaguchi:2002pa}
S.~Yamaguchi, ``{Holographic RG flow on the defect and g theorem},''
  \href{http://dx.doi.org/10.1088/1126-6708/2002/10/002}{{\em JHEP} {\bfseries
  10} (2002) 002}, \href{http://arxiv.org/abs/hep-th/0207171}{{\ttfamily
  arXiv:hep-th/0207171}}.

\bibitem{Azeyanagi:2007qj}
T.~Azeyanagi, A.~Karch, T.~Takayanagi, and E.~G. Thompson, ``{Holographic
  calculation of boundary entropy},''
  \href{http://dx.doi.org/10.1088/1126-6708/2008/03/054}{{\em JHEP} {\bfseries
  03} (2008) 054}, \href{http://arxiv.org/abs/0712.1850}{{\ttfamily
  arXiv:0712.1850 [hep-th]}}.

\bibitem{Estes:2014hka}
J.~Estes, K.~Jensen, A.~O'Bannon, E.~Tsatis, and T.~Wrase, ``{On Holographic
  Defect Entropy},'' \href{http://dx.doi.org/10.1007/JHEP05(2014)084}{{\em
  JHEP} {\bfseries 05} (2014) 084},
  \href{http://arxiv.org/abs/1403.6475}{{\ttfamily arXiv:1403.6475 [hep-th]}}.

\bibitem{Andrei:2018die}
N.~Andrei {\em et~al.}, ``{Boundary and Defect CFT: Open Problems and
  Applications},'' \href{http://dx.doi.org/10.1088/1751-8121/abb0fe}{{\em J.
  Phys. A} {\bfseries 53} no.~45, (2020) 453002},
  \href{http://arxiv.org/abs/1810.05697}{{\ttfamily arXiv:1810.05697
  [hep-th]}}.

\bibitem{Kobayashi:2018lil}
N.~Kobayashi, T.~Nishioka, Y.~Sato, and K.~Watanabe, ``{Towards a $C$-theorem
  in defect CFT},'' \href{http://dx.doi.org/10.1007/JHEP01(2019)039}{{\em JHEP}
  {\bfseries 01} (2019) 039}, \href{http://arxiv.org/abs/1810.06995}{{\ttfamily
  arXiv:1810.06995 [hep-th]}}.

\bibitem{Casini:2018nym}
H.~Casini, I.~Salazar~Landea, and G.~Torroba, ``{Irreversibility in quantum
  field theories with boundaries},''
  \href{http://dx.doi.org/10.1007/JHEP04(2019)166}{{\em JHEP} {\bfseries 04}
  (2019) 166}, \href{http://arxiv.org/abs/1812.08183}{{\ttfamily
  arXiv:1812.08183 [hep-th]}}.

\bibitem{Lauria:2020emq}
E.~Lauria, P.~Liendo, B.~C. Van~Rees, and X.~Zhao, ``{Line and surface defects
  for the free scalar field},''
  \href{http://dx.doi.org/10.1007/JHEP01(2021)060}{{\em JHEP} {\bfseries 01}
  (2021) 060}, \href{http://arxiv.org/abs/2005.02413}{{\ttfamily
  arXiv:2005.02413 [hep-th]}}.

\bibitem{Giombi:2020rmc}
S.~Giombi and H.~Khanchandani, ``{CFT in AdS and boundary RG flows},''
  \href{http://dx.doi.org/10.1007/JHEP11(2020)118}{{\em JHEP} {\bfseries 11}
  (2020) 118}, \href{http://arxiv.org/abs/2007.04955}{{\ttfamily
  arXiv:2007.04955 [hep-th]}}.

\bibitem{Wang:2020xkc}
Y.~Wang, ``{Surface defect, anomalies and b-extremization},''
  \href{http://dx.doi.org/10.1007/JHEP11(2021)122}{{\em JHEP} {\bfseries 11}
  (2021) 122}, \href{http://arxiv.org/abs/2012.06574}{{\ttfamily
  arXiv:2012.06574 [hep-th]}}.

\bibitem{Nishioka:2021uef}
T.~Nishioka and Y.~Sato, ``{Free energy and defect $C$-theorem in free scalar
  theory},'' \href{http://dx.doi.org/10.1007/JHEP05(2021)074}{{\em JHEP}
  {\bfseries 05} (2021) 074}, \href{http://arxiv.org/abs/2101.02399}{{\ttfamily
  arXiv:2101.02399 [hep-th]}}.

\bibitem{Sato:2021eqo}
Y.~Sato, ``{Free energy and defect $C$-theorem in free fermion},''
  \href{http://dx.doi.org/10.1007/JHEP05(2021)202}{{\em JHEP} {\bfseries 05}
  (2021) 202}, \href{http://arxiv.org/abs/2102.11468}{{\ttfamily
  arXiv:2102.11468 [hep-th]}}.

\bibitem{CarrenoBolla:2023vrv}
I.~Carre\~no Bolla, D.~Rodriguez-Gomez, and J.~G. Russo, ``{Defects, rigid
  holography, and C-theorems},''
  \href{http://dx.doi.org/10.1103/PhysRevD.108.L041701}{{\em Phys. Rev. D}
  {\bfseries 108} no.~4, (2023) L041701},
  \href{http://arxiv.org/abs/2306.11796}{{\ttfamily arXiv:2306.11796
  [hep-th]}}.

\bibitem{Padayasi:2021sik}
J.~Padayasi, A.~Krishnan, M.~A. Metlitski, I.~A. Gruzberg, and M.~Meineri,
  ``{The extraordinary boundary transition in the 3d O(N) model via conformal
  bootstrap},'' \href{http://dx.doi.org/10.21468/SciPostPhys.12.6.190}{{\em
  SciPost Phys.} {\bfseries 12} no.~6, (2022) 190},
  \href{http://arxiv.org/abs/2111.03071}{{\ttfamily arXiv:2111.03071
  [cond-mat.stat-mech]}}.

\bibitem{Rodriguez-Gomez:2022gbz}
D.~Rodriguez-Gomez, ``{A scaling limit for line and surface defects},''
  \href{http://dx.doi.org/10.1007/JHEP06(2022)071}{{\em JHEP} {\bfseries 06}
  (2022) 071}, \href{http://arxiv.org/abs/2202.03471}{{\ttfamily
  arXiv:2202.03471 [hep-th]}}.

\bibitem{Cuomo:2021kfm}
G.~Cuomo, Z.~Komargodski, and M.~Mezei, ``{Localized magnetic field in the O(N)
  model},'' \href{http://dx.doi.org/10.1007/JHEP02(2022)134}{{\em JHEP}
  {\bfseries 02} (2022) 134}, \href{http://arxiv.org/abs/2112.10634}{{\ttfamily
  arXiv:2112.10634 [hep-th]}}.

\bibitem{Rodriguez-Gomez:2022gif}
D.~Rodriguez-Gomez and J.~G. Russo, ``{Defects in scalar field theories, RG
  flows and dimensional disentangling},''
  \href{http://dx.doi.org/10.1007/JHEP11(2022)167}{{\em JHEP} {\bfseries 11}
  (2022) 167}, \href{http://arxiv.org/abs/2209.00663}{{\ttfamily
  arXiv:2209.00663 [hep-th]}}.

\bibitem{Castiglioni:2022yes}
L.~Castiglioni, S.~Penati, M.~Tenser, and D.~Trancanelli, ``{Interpolating
  Wilson loops and enriched RG flows},''
  \href{http://arxiv.org/abs/2211.16501}{{\ttfamily arXiv:2211.16501
  [hep-th]}}.

\bibitem{Krishnan:2023cff}
A.~Krishnan and M.~A. Metlitski, ``{A plane defect in the 3d O$(N)$ model},''
  \href{http://arxiv.org/abs/2301.05728}{{\ttfamily arXiv:2301.05728
  [cond-mat.str-el]}}.

\bibitem{Cuomo:2023qvp}
G.~Cuomo and S.~Zhang, ``{Spontaneous symmetry breaking on surface defects},''
  \href{http://dx.doi.org/10.1007/JHEP03(2024)022}{{\em JHEP} {\bfseries 03}
  (2024) 022}, \href{http://arxiv.org/abs/2306.00085}{{\ttfamily
  arXiv:2306.00085 [hep-th]}}.

\bibitem{Drukker:2023jxp}
N.~Drukker, O.~Shahpo, and M.~Tr\'epanier, ``{Quantum holographic surface
  anomalies},'' \href{http://dx.doi.org/10.1088/1751-8121/ad2296}{{\em J. Phys.
  A} {\bfseries 57} no.~8, (2024) 085402},
  \href{http://arxiv.org/abs/2311.14797}{{\ttfamily arXiv:2311.14797
  [hep-th]}}.

\bibitem{Shimamori:2024yms}
S.~Shimamori, ``{Conformal field theory with composite defect},''
  \href{http://arxiv.org/abs/2404.08411}{{\ttfamily arXiv:2404.08411
  [hep-th]}}.

\bibitem{Friedan:2003yc}
D.~Friedan and A.~Konechny, ``{On the boundary entropy of one-dimensional
  quantum systems at low temperature},''
  \href{http://dx.doi.org/10.1103/PhysRevLett.93.030402}{{\em Phys. Rev. Lett.}
  {\bfseries 93} (2004) 030402},
  \href{http://arxiv.org/abs/hep-th/0312197}{{\ttfamily arXiv:hep-th/0312197}}.

\bibitem{Casini:2016fgb}
H.~Casini, I.~Salazar~Landea, and G.~Torroba, ``{The g-theorem and quantum
  information theory},'' \href{http://dx.doi.org/10.1007/JHEP10(2016)140}{{\em
  JHEP} {\bfseries 10} (2016) 140},
  \href{http://arxiv.org/abs/1607.00390}{{\ttfamily arXiv:1607.00390
  [hep-th]}}.

\bibitem{Cuomo:2021rkm}
G.~Cuomo, Z.~Komargodski, and A.~Raviv-Moshe, ``{Renormalization Group Flows on
  Line Defects},'' \href{http://dx.doi.org/10.1103/PhysRevLett.128.021603}{{\em
  Phys. Rev. Lett.} {\bfseries 128} no.~2, (2022) 021603},
  \href{http://arxiv.org/abs/2108.01117}{{\ttfamily arXiv:2108.01117
  [hep-th]}}.

\bibitem{Casini:2022bsu}
H.~Casini, I.~Salazar~Landea, and G.~Torroba, ``{The entropic $g$-theorem in
  general spacetime dimension},''
  \href{http://arxiv.org/abs/2212.10575}{{\ttfamily arXiv:2212.10575
  [hep-th]}}.

\bibitem{Jensen:2015swa}
K.~Jensen and A.~O'Bannon, ``{Constraint on Defect and Boundary Renormalization
  Group Flows},'' \href{http://dx.doi.org/10.1103/PhysRevLett.116.091601}{{\em
  Phys. Rev. Lett.} {\bfseries 116} no.~9, (2016) 091601},
  \href{http://arxiv.org/abs/1509.02160}{{\ttfamily arXiv:1509.02160
  [hep-th]}}.

\bibitem{Wang:2021mdq}
Y.~Wang, ``{Defect a-theorem and a-maximization},''
  \href{http://dx.doi.org/10.1007/JHEP02(2022)061}{{\em JHEP} {\bfseries 02}
  (2022) 061}, \href{http://arxiv.org/abs/2101.12648}{{\ttfamily
  arXiv:2101.12648 [hep-th]}}.

\bibitem{Shachar:2022fqk}
T.~Shachar, R.~Sinha, and M.~Smolkin, ``{RG flows on two-dimensional spherical
  defects},'' \href{http://dx.doi.org/10.21468/SciPostPhys.15.6.240}{{\em
  SciPost Phys.} {\bfseries 15} no.~6, (2023) 240},
  \href{http://arxiv.org/abs/2212.08081}{{\ttfamily arXiv:2212.08081
  [hep-th]}}.

\bibitem{Green:2007wr}
D.~R. Green, M.~Mulligan, and D.~Starr, ``{Boundary Entropy Can Increase Under
  Bulk RG Flow},''
  \href{http://dx.doi.org/10.1016/j.nuclphysb.2008.01.010}{{\em Nucl. Phys. B}
  {\bfseries 798} (2008) 491--504},
  \href{http://arxiv.org/abs/0710.4348}{{\ttfamily arXiv:0710.4348 [hep-th]}}.

\bibitem{Sato:2020upl}
Y.~Sato, ``{Boundary entropy under ambient RG flow in the AdS/BCFT model},''
  \href{http://dx.doi.org/10.1103/PhysRevD.101.126004}{{\em Phys. Rev. D}
  {\bfseries 101} no.~12, (2020) 126004},
  \href{http://arxiv.org/abs/2004.04929}{{\ttfamily arXiv:2004.04929
  [hep-th]}}.

\bibitem{Trepanier2023}
M.~Trépanier, ``Surface defects in the o(n) model,''
  \href{http://dx.doi.org/10.1007/JHEP09(2023)074}{{\em Journal of High Energy
  Physics} {\bfseries 2023} no.~9, (2023) 74}.
  \url{https://doi.org/10.1007/JHEP09(2023)074}.

\bibitem{Giombi2023}
S.~Giombi and B.~Liu, ``Notes on a surface defect in the o(n) model,''
  \href{http://dx.doi.org/10.1007/JHEP12(2023)004}{{\em Journal of High Energy
  Physics} {\bfseries 2023} no.~12, (2023) 4}.
  \url{https://doi.org/10.1007/JHEP12(2023)004}.

\bibitem{RavivMoshe2023}
A.~Raviv-Moshe and S.~Zhong, ``Phases of surface defects in scalar field
  theories,'' \href{http://dx.doi.org/10.1007/JHEP08(2023)143}{{\em Journal of
  High Energy Physics} {\bfseries 2023} no.~8, (2023) 143}.
  \url{https://doi.org/10.1007/JHEP08(2023)143}.

\bibitem{Casini:2023kyj}
H.~Casini, I.~Salazar~Landea, and G.~Torroba, ``{Irreversibility, QNEC, and
  defects},'' \href{http://dx.doi.org/10.1007/JHEP07(2023)004}{{\em JHEP}
  {\bfseries 07} (2023) 004}, \href{http://arxiv.org/abs/2303.16935}{{\ttfamily
  arXiv:2303.16935 [hep-th]}}.

\bibitem{Yuan:2022oeo}
M.-K. Yuan and Y.~Zhou, ``{Defect localized entropy: Renormalization group and
  holography},'' \href{http://dx.doi.org/10.1016/j.nuclphysb.2023.116301}{{\em
  Nucl. Phys. B} {\bfseries 994} (2023) 116301},
  \href{http://arxiv.org/abs/2209.08835}{{\ttfamily arXiv:2209.08835
  [hep-th]}}.

\bibitem{Yuan:2023oni}
M.-K. Yuan and Y.~Zhou, ``{R\'enyi entropy with surface defects in six
  dimensions},'' \href{http://dx.doi.org/10.1007/JHEP03(2024)031}{{\em JHEP}
  {\bfseries 03} (2024) 031}, \href{http://arxiv.org/abs/2310.02096}{{\ttfamily
  arXiv:2310.02096 [hep-th]}}.

\bibitem{Harper:2024aku}
J.~Harper, H.~Kanda, T.~Takayanagi, and K.~Tasuki, ``{The $g$-theorem from
  Strong Subadditivity},'' \href{http://arxiv.org/abs/2403.19934}{{\ttfamily
  arXiv:2403.19934 [hep-th]}}.

\bibitem{Klebanov2011}
I.~R. Klebanov, S.~S. Pufu, and B.~R. Safdi, ``F-theorem without
  supersymmetry,'' \href{http://dx.doi.org/10.1007/JHEP10(2011)038}{{\em
  Journal of High Energy Physics} {\bfseries 2011} no.~10, (2011) 38}.
  \url{https://doi.org/10.1007/JHEP10(2011)038}.

\bibitem{Goykhman2021}
M.~Goykhman, V.~Rosenhaus, and M.~Smolkin, ``The background field method and
  critical vector models,''
  \href{http://dx.doi.org/10.1007/JHEP02(2021)074}{{\em Journal of High Energy
  Physics} {\bfseries 2021} no.~2, (2021) 74}.
  \url{https://doi.org/10.1007/JHEP02(2021)074}.

\bibitem{ZinnJustin1991}
J.~Zinn-Justin, ``Four-fermion interaction near four dimensions,''
  \href{http://dx.doi.org/https://doi.org/10.1016/0550-3213(91)90043-W}{{\em
  Nuclear Physics B} {\bfseries 367} no.~1, (1991) 105--122}.
  \url{https://www.sciencedirect.com/science/article/pii/055032139190043W}.

\bibitem{Moshe2003}
M.~Moshe and J.~Zinn-Justin, ``Quantum field theory in the large n limit: a
  review,''
  \href{http://dx.doi.org/https://doi.org/10.1016/S0370-1573(03)00263-1}{{\em
  Physics Reports} {\bfseries 385} no.~3, (2003) 69--228}.
  \url{https://www.sciencedirect.com/science/article/pii/S0370157303002631}.

\end{thebibliography}\endgroup
	
\end{document}